\documentclass[%
 aip,
 jcp,%
 amsmath,amssymb,
 reprint,%
]{revtex4-1}

\usepackage{graphicx}
\usepackage{dcolumn}
\usepackage{bm}

\begin{document}

\title{Equation of state and shock compression of warm dense sodium –-- a first-principles study}

\author{Shuai Zhang}
\email{shuai.zhang01@berkeley.edu}
\author{Kevin P. Driver}
\author{Fran\c{c}ois Soubiran}
\affiliation{Department of Earth and Planetary Science, University of California, Berkeley, California 94720, USA}
\author{Burkhard Militzer}
\email{militzer@berkeley.edu}
\affiliation{Department of Earth and Planetary Science, University of California, Berkeley, California 94720, USA}
\affiliation{Department of Astronomy, University of California, Berkeley, California 94720, USA}

\date{\today}

\begin{abstract}
{As one of the simple alkali metals, sodium has been of fundamental
interest for shock physics experiments, but knowledge of its equation
of state (EOS) in hot, dense regimes is not well known. By combining
path integral Monte Carlo (PIMC) results for partially-ionized
states {[B. Militzer and K. P. Driver, Phys. Rev. Lett. \textbf{115}, 176403 (2015)]}
at high temperatures and density functional
theory molecular dynamics (DFT-MD) results at lower temperatures, we
have constructed a coherent equation of state for sodium over a
wide density-temperature range of $1.93-11.60$ g/cm$^{3}$ and
$10^3-1.29\times10^8$~K. We find that a localized, Hartree-Fock nodal
structure in PIMC yields pressures and internal energies
that are consistent with DFT-MD at intermediate temperatures of
$2\times10^6$ K. Since PIMC and DFT-MD provide a first-principles treatment
of electron shell and excitation effects, we are able to identify two
compression maxima in the shock Hugoniot curve corresponding to
$K$-shell and $L$-shell ionization. Our Hugoniot curves provide a
benchmark for widely-used EOS models, SESAME, LEOS, and
Purgatorio. Due to the low ambient density, sodium has
an unusually high first compression maximum
along the shock Hugoniot curve.  At beyond 10$^7$~K,
we show that the radiation effect leads to very
high compression along the Hugoniot curve, surpassing
relativistic corrections, and observe an increasing
deviation of the shock and particle velocities from a linear
relation. We also compute the temperature-density dependence 
of thermal and pressure ionization processes.}
\end{abstract}



\maketitle

\section{Introduction}   

The behavior of sodium and other alkali metals at extreme conditions
has generated intense scientific interest over many decades as
experimental and theoretical technology has evolved to facilitate
studies of increasingly atypical states. At ambient conditions, sodium
has long been known as a prototypical, simple, free-electron
metal~\cite{Wigner1933} with a high-symmetry, bulk-centered cubic (bcc) structure. Recent
interest in sodium has been driven by the the ability to carefully
probe exotic high-pressure states~\cite{Neaton2001}, made possible by
improvements in static-compression~\cite{Eremets1996,Dubrovinsky2012}
and single-crystal experimental techniques~\cite{Gregoryanz2008}. A
number of experimental and theoretical works have revealed that sodium
exhibits anomalous
structural~\cite{Aleksandrov1982,Fritz1984,Hanfland2002,Raty2007,McMahon2007,Gregoryanz2008,Rousseau2011},
electronic and
optical~\cite{Raty2007,Ma2009,Lazicki2009,Gatti2010,Mao20122011,Marques2011,Loa2011,Pozzo2011,IbaNez-Azpiroz2014,Miao2015,Naumov2015},
and
melting~\cite{Gregoryanz2005,Koci2008,Aki2008,Maksimov2011,Marques2011,Eshet2012,Gonzalez2015}
behavior at high pressure. Among the structural studies,
high-pressure X-ray diffraction experiments have explored sodium up
to pressures and temperatures of 150 GPa and 1000 K, observing a
complex series of phase
transitions~\cite{Hanfland2002,McMahon2007,Gregoryanz2008} from the
bcc phase to at least seven lower-symmetry phases. Sodium also
displays an anomolous melting curve maximum~\cite{Gregoryanz2005} in
this pressure range as the liquid undergoes structural electronic
changes, becoming more dense than the solid over a large pressure
range of $\sim$60 GPa.  At even higher pressures, beyond 200 GPa,
sodium undergoes a metal-insulator transition~\cite{Ma2009} by forming
an electride~\cite{Miao2015}.
\begin{figure}[!htbp]
\centering\includegraphics[width=0.5\textwidth]{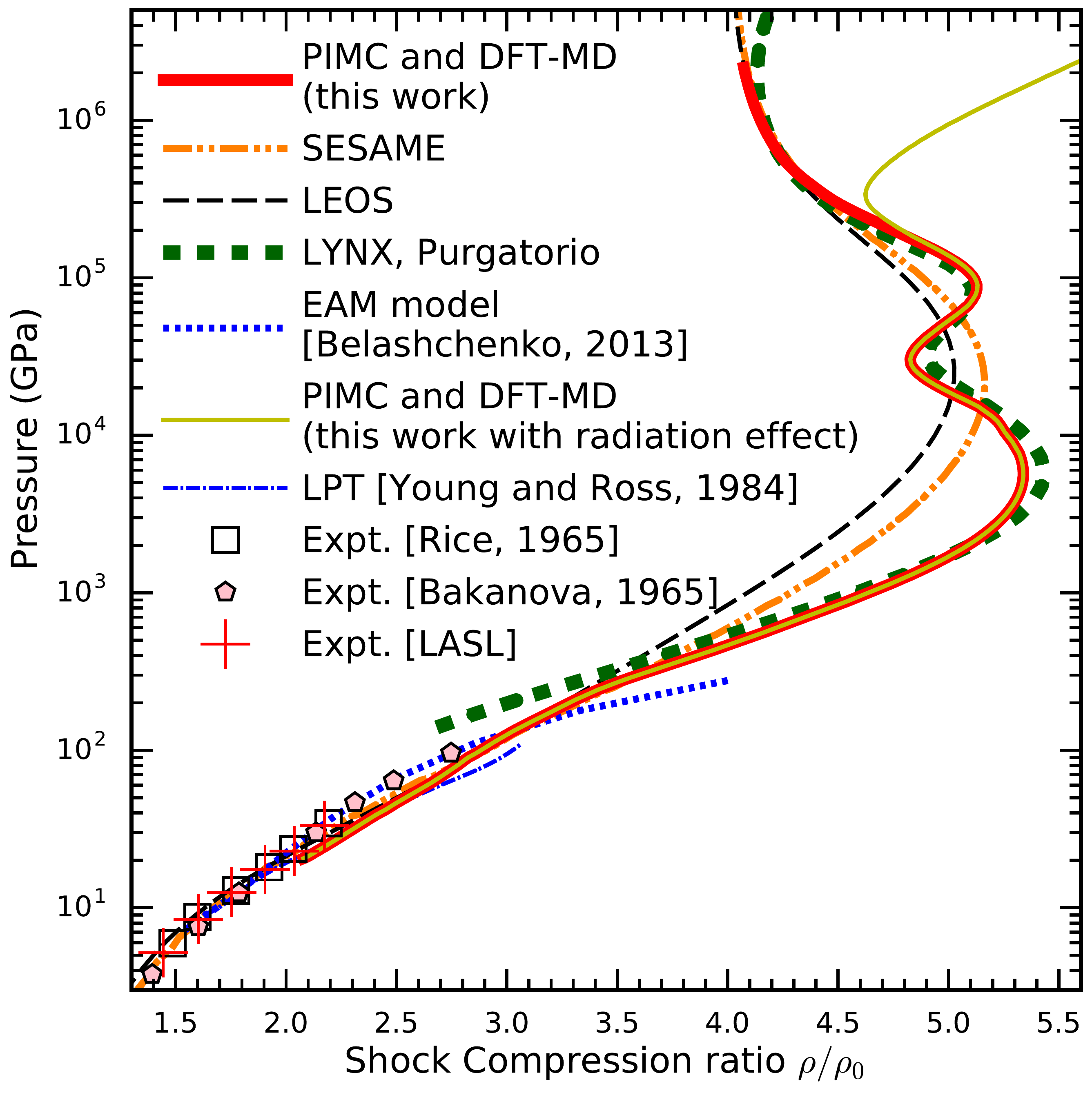}
\caption{\label{fig0} Principal Hugoniot of Na obtained in this work
  in comparison with that considering the photon radiation
  correction and those from SESAME, LEOS, and Purgatorio (Lynx)
  models.}
\end{figure}
\emph{Ab initio} random
structure searches predict that sodium will ultimately become a
reentrant metal at a pressure of 15.5 TPa at 0 K~\cite{Li2015}.

Analogous to the impact that improved static-compression technology
has had for alkali metals in the condensed matter regime, we can expect
future dynamic compression research to routinely probe the
gigabar (Gbar)~\cite{Swift2012,Kritcher2016} pressures and provide new
data on the behavior of warm dense matter (WDM). Shock
measurements of hot, dense alkali metals are expected to be
an important component of solving problems in astrophysics, planetary
physics, stockpile stewardship, and inertial fusion.  Thus far, only
relatively low pressure ($P < 228$~GPa~\cite{Golyshev2011}, $T <
8168$~K) shock experiments have been performed on sodium. Early
dynamic compression experiments were motivated by compressibility of
metals and the simplicity of ambient sodium as a material for the
study of fundamental shock physics
behavior~\cite{Rice1965,Bakanova1965,MarshLASL1980}. More recent shock
experiments have been motivated by the prospect of observing the
metal-insulator transition~\cite{Golyshev2011}.

Complementary to the shock experiments, there have been a number of
theoretical works aiming at calculating the equation
of state (EOS) and shock
Hugoniot curve using either analytic
approaches~\cite{Ehrenfeld1966analyticHug,Hickey1967comment} or
semi-classical free-energy
models~\cite{Pastine1967,Pastine1968,Khishchenko2008},
fluid-variational theory~\cite{Ross1980}, local pseudopotential
theory~\cite{Young1984}, or EAM
potentials~\cite{Belashchenko2009,Belashchenko2013a,Belashchenko2013b}.
While each of these methods agrees reasonably well with the available
low-pressure shock Hugoniot data ($P < 100$ GPa), none possess the
rigor that is necessary to treat the strong coupling, quantum
degeneracy, and ionization physics associated with hot, dense plasma
regimes, that Gbar-shock experiments will access.

Development of rigorous, first-principles frameworks, such as 
path integral Monte Carlo
(PIMC)~\cite{PhysRevLett.73.2145,Driver2012,Militzer2015} or
density functional theory molecular dynamics (DFT-MD)~\cite{PhysRevB.94.094109,PhysRevE.86.026405,PhysRevB.94.205115,ZhangExtendedDFT2016}, for computing the
EOS and behavior of materials in extreme pressure
and temperature conditions is needed as Gbar-range 
shock experiments are emerging and data need to be
interpreted. Recently, we have been developing PIMC for simulating
heavier
elements~\cite{PhysRevE.63.066404,Militzer2009,Driver2012,Benedict2014C,Driver2015Neon,Militzer2015,Driver2015Oxygen,Driver2016Nitrogen,PhysRevB.94.094109,Zhang2016b},
which is efficient at high temperatures, and can be used to complement
DFT-MD calculations that are efficient at comparatively low
temperatures. Combined data from PIMC and DFT-MD provide a coherent
EOS over a wide density-temperature range that spans the condensed
matter, WDM, and plasma regimes. In the current work, we use PIMC and
DFT-MD to compute the EOS, shock compression behavior, and plasma
structure evolution of sodium across a much larger density-temperature
range than has ever been studied in experiments or in theory.  We
provide our EOS table and shock Hugoniot curve as a theoretical benchmark in
comparison with widely-used EOS database models and experiments.

The paper is organized as follows: Section~\ref{method} discusses details of
our PIMC and DFT-MD simulation methods. In Sec.~\ref{results}, we discuss
several results: (1) comparison of the performance of different
choices for PIMC nodal surfaces, (2) the internal energies and
pressures, (3) shock Hugoniot curves, and 
(4) the density-temperature evolution of ionization
processes. Finally, in Sec.~\ref{conclusion}, we conclude.

\section{Simulation methods}\label{method}
A reliable theoretical scheme for simulating materials across a wide
range of density and temperature conditions naturally involves treating physics
appropriately in different regimes.  In the limit of high temperature
(above $10^8$ K), materials tend to behave as weakly-interacting
plasmas because of strong thermal ionization. Under such conditions,
the ideal Fermi gas model and the Debye-H\"{u}ckel
theory~\cite{DebyeHuckel} are good approximations. At low to
intermediate temperatures ($T<10^6$ K), standard, orbital-based,
Kohn-Sham DFT-MD is a
suitable option to derive the EOS because it fully accounts the
bonding effects and bound states within certain approximations of the
exchange-correlation functional. However, the need to explicitly
compute all partially occupied electronic orbitals causes DFT-MD
to become computationally intractable beyond temperatures of roughly
$10^6$ K.  Recent work on orbital-free
DFT~\cite{Sjostrom2014,PhysRevB.94.094109,Karasiev2016,PhysRevB.94.205115} and
extended DFT-MD with free electron approximation for high
energies~\cite{ZhangExtendedDFT2016} have made progress toward
overcoming this difficulty, but their general applicability and
accuracy needs to be further examined.


PIMC simulates all quantum, many-body
exchange and correlations effects and provides the most natural
formulation to compute accurate EOSs at high temperatures ($T>10^6$
K).  In PIMC, the thermal density matrix is expressed as Feynman path
integral~\cite{Feynman1953}, which treats electrons and nuclei as
quantum paths that are cyclic in the imaginary time
$0\le t\le\beta=1/k_\text{B}T$, where $k_\text{B}$ is the Boltzmann
constant. Therefore, PIMC becomes increasingly efficient at higher
temperatures as paths become shorter and more classical in
nature. However, application of PIMC to study real materials other
than hydrogen
\cite{PhysRevLett.73.2145,PhysRevLett.76.1240,PhysRevE.63.066404,PhysRevLett.87.275502,PhysRevLett.104.235003,PhysRevLett.85.1890,PhysRevB.84.224109,CTPP:CTPP2150390137,Militzer20062136},
helium \cite{Militzer2009,PhysRevLett.97.175501}, hydrogen-helium
mixtures \cite{Militzer2005}, and one-component plasmas
\cite{PhysRevB.71.134303,PhysRevLett.92.021101}, is difficult because
of the complex fermion sign problem, nonlocal pseudopotentials, and
complex nodal structures~\cite{Ceperley1991}.  

The sign problem in fermionic PIMC simulations is
usually addressed with the fixed-node approximation
\cite{Ceperley1991} that restricts paths to positive regions of a
trial density matrix, $\rho_T(\textbf{R},\textbf{R}_t; t)>0$.  The
restricted path integral reads,
\begin{equation}
\rho_F(\textbf{R}, \textbf{R}' ;\beta) =
\frac{1}{N!}\sum_\mathcal{P}(-1)^\mathcal{P} \int_{\textbf{R}\rightarrow\mathcal{P}\textbf{R}', \;\rho_T>0}d\textbf{R}_t\;e^{-S[\textbf{R}_t]},
\end{equation}
where $\mathcal{P}$ denotes permutations of identical particles, and
$S$ is the action that weights the paths. The most common
approximation to the trial density matrix is a Slater determinant of
single-particle density matrices,
\begin{equation}
\rho_T(\textbf{R},\textbf{R}';\beta)=\left|\left| \rho^{[1]}(\textbf{\textit{r}}_{i},\textbf{\textit{r}}'_{j};\beta) \right|\right|_{ij},
\end{equation}
in combination with the free-particle (FP) density matrix,
\begin{equation}
\rho^{[1]}_0(\textbf{\textit{r}},\textbf{\textit{r}}';\beta) = \sum_\textbf{\textit{k}} e^{-\beta E_\textbf{\textit{k}}}\Psi_\textbf{\textit{k}}(\textbf{\textit{r}})\Psi_\textbf{\textit{k}}^*(\textbf{\textit{r}}'),
\end{equation}
which represents a sum over all plane waves 
$\Psi_\textbf{\textit{k}}(\textbf{\textit{r}})$ 
with energy $E_\textbf{\textit{k}}$.

PIMC with FP nodes gives exact results in the limit of high
temperature \cite{Militzer2009}. Developments of this method have allowed for
remarkable progress. Using FP nodes, we have successfully obtained the
EOS of several heavy elements (C \cite{Driver2012}, N
\cite{Driver2016Nitrogen}, O \cite{Driver2015Oxygen}, Ne
\cite{Driver2015Neon}) and compounds (H$_2$O \cite{Driver2012}).  The
combined results of PIMC and DFT-MD have bridged the WDM gap between
DFT-MD and the high temperature limit and provided consistent sets of
coherent EOS from first principles.

Several attempts have been made to go beyond FP nodes in PIMC
simulations \cite{Khairallah2011,PhysRevLett.85.1890}. In recent
work~\cite{Militzer2015}, we show that the applicability range of
PIMC simulations can extend to lower temperatures when a number of
$n_s$ atomic orbitals at each ion $I$ are added to the FP nodes,
\begin{equation}  \label{pollock}
\begin{split}
  \rho^{[1]}(\textbf{\textit{r}},\textbf{\textit{r}}',\beta) & =  \rho^{[1]}_0(\textbf{\textit{r}},\textbf{\textit{r}}';\beta) + \\
  & \sum_{I=1}^{N} \sum_{s=0}^{n_s} e^{-\beta E_s} \Psi_s(\textbf{\textit{r}}-\textbf{\textit{R}}_I) \Psi_s^*(\textbf{\textit{r}}'-\textbf{\textit{R}}_I)\;.
\end{split}
\end{equation}
Our results for a single silicon atom in
periodic boundary condition showed that nodes derived from
Hartree-Fock (HF) orbitals yield highly accurate predictions for the
pressure and the internal energy at much lower temperature than is
accessible with FP nodes. The combined results using this PIMC method
and DFT-MD provided a coherent EOS for dense silicon plasmas over a
wide density-temperature grid (2.3-18.6 g/cm$^{3}$,
$5\times10^5$-$1.3\times10^8$ K).  In this work, we will also
investigate the effects of various nodal surfaces in PIMC calculations and
show that the localized, Hartree-Fock orbitals yield accurate
pressures and internal energies for sodium.

Our PIMC simulations are performed within the fixed node
approximation~\cite{Ceperley1991} and based on the {\footnotesize
  CUPID} code \cite{militzerphd}.  Similar to the PIMC simulations of
Si \cite{Militzer2015} and recently of Na at one density
\cite{Zhang2016b}, we treat the nuclei classically because of the high
temperatures considered here.  Electronic Coulomb interactions are
introduced via pair density matrices \cite{pdm}. 

DFT-MD simulations use the Vienna \textit{Ab initio} Simulation Package
({\footnotesize VASP}) \cite{kresse96b} and implement exchange-correlation functionals
within the local density approximation (LDA)
\cite{Perdew81,Ceperley1980}. 
{The effect of using other forms of exchange-correlation functionals is small,
in comparison to the effects of thermal excitations,
as will be disscussed in Sec.~\ref{IIa}.}
We use a projected augmented wave (PAW)
pseudopotential \cite{Blochl1994} with a frozen, 1s$^2$ electrons core
and a small core radius of 1.45 Bohr,
{which is the hardest one available for Na in \scriptsize{VASP}~\cite{comment1}.}

{Since the width of the electronic bands is small
compared to the thermal excitation energies,
we use the $\Gamma$ point for sampling the Brillouin zone.
We choose  plane wave basis with a large cutoff of 4000 eV to account for
the electrons that are excited to the very high-energy states. }
{We use a time step of $0.2$ fs and perform DFT-MD simulations typically for over
2000 steps. 
A large number of temperature, energy, and pressure 
fluctuations are observed,
which indicate the simulations have reached equilibrium.
We discard 20\% of the trajectories at the beginning and
average the internal energy and pressure
to derive the EOS.}
 In order to put the PAW-LDA pseudopotential energies
on the same scale as all-electron calculations, we shifted all of our
{\footnotesize VASP} DFT-MD energies by -161.3386 Ha/atom. This shift was determined
by performing isolated, all-electron atomic calculations with the
{\footnotesize OPIUM} code~\cite{opium} and corresponding 
isolated-atom calculations in {\footnotesize VASP}.

\begin{figure}
\centering\includegraphics[width=0.5\textwidth]{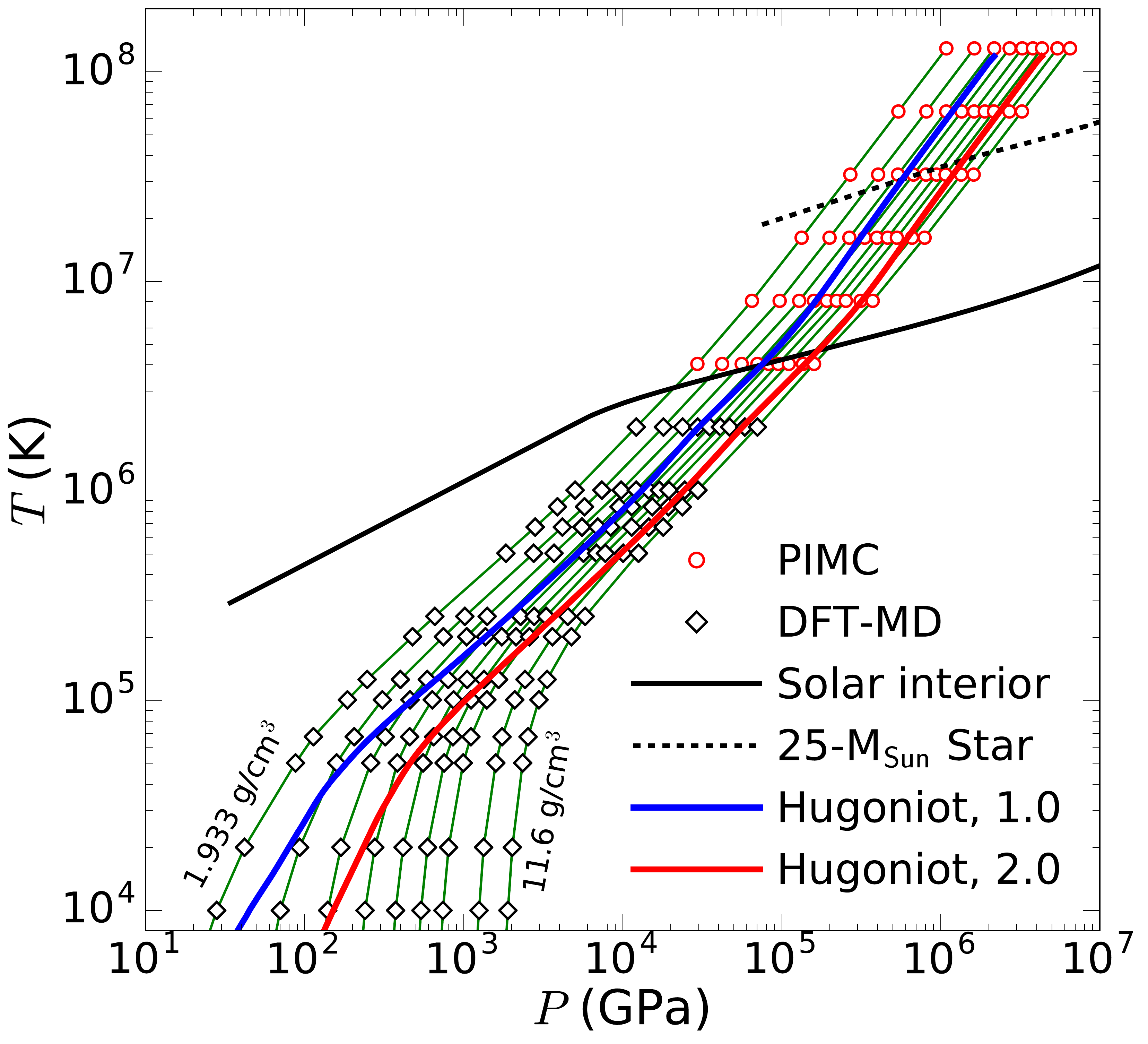}
\caption{\label{fig1} Nine isochores 
(corresponding to 2-12 $\rho_\text{ambient}$) 
in a temperature-pressure plot, 
along with the interior profiles of the Sun
\cite{PhysRevLett.92.121301} and a star 
at the end of its helium burning \cite{CGpt}. 
The $P-T$ conditions of our
PIMC and DFT-MD simulations and two Hugoniot curves
with initial densities $\rho_0$ 
of 1.0 and 2.0 times the ambient density
are also shown.}
\end{figure}

Most of our simulations
{start from a bcc cell} with 8 atoms. 
{In comparison with using a 54-atom cell,
DFT-MD simulations with the 8-atom cell is sufficient in providing
well converged internal energies and pressures down to temperatures
 below 5.0$\times10^4$ K. For example, the difference is 0.1 eV/atom
in internal energy and 0.3 GPa in pressure between simulations using an
8- and a 54-atom cells, at 3.87 g/cm$^3$ and 5.05$\times10^4$ K.
At lower temperatures, the system could tend to freeze
and is subject to larger finite-size errors. In order to maximally
eliminate this effect, we use a larger, 54-atom cell at these temperatures
up to 2.0$\times10^5$ K, as long as the computational cost is affordable.}
We simulate data along nine isochores corresponding to 2-, 3-,
4-, 5-, 6-, 7-, 8-, 10-, and 12-fold compression of ambient density
$\rho_\text{ambient}$=0.96663 g/cm$^{3}$. For each density, we study
the temperature range from 10$^3$ to 1.29$\times10^8$ K,
relevant to the regime of WDM and stellar interiors (Figs.~\ref{fig0}-\ref{fig1}).

\section{Results}\label{results}

\subsection{Comparison of PIMC methods}\label{IIa}

\begin{figure}
\centering\includegraphics[width=0.5\textwidth]{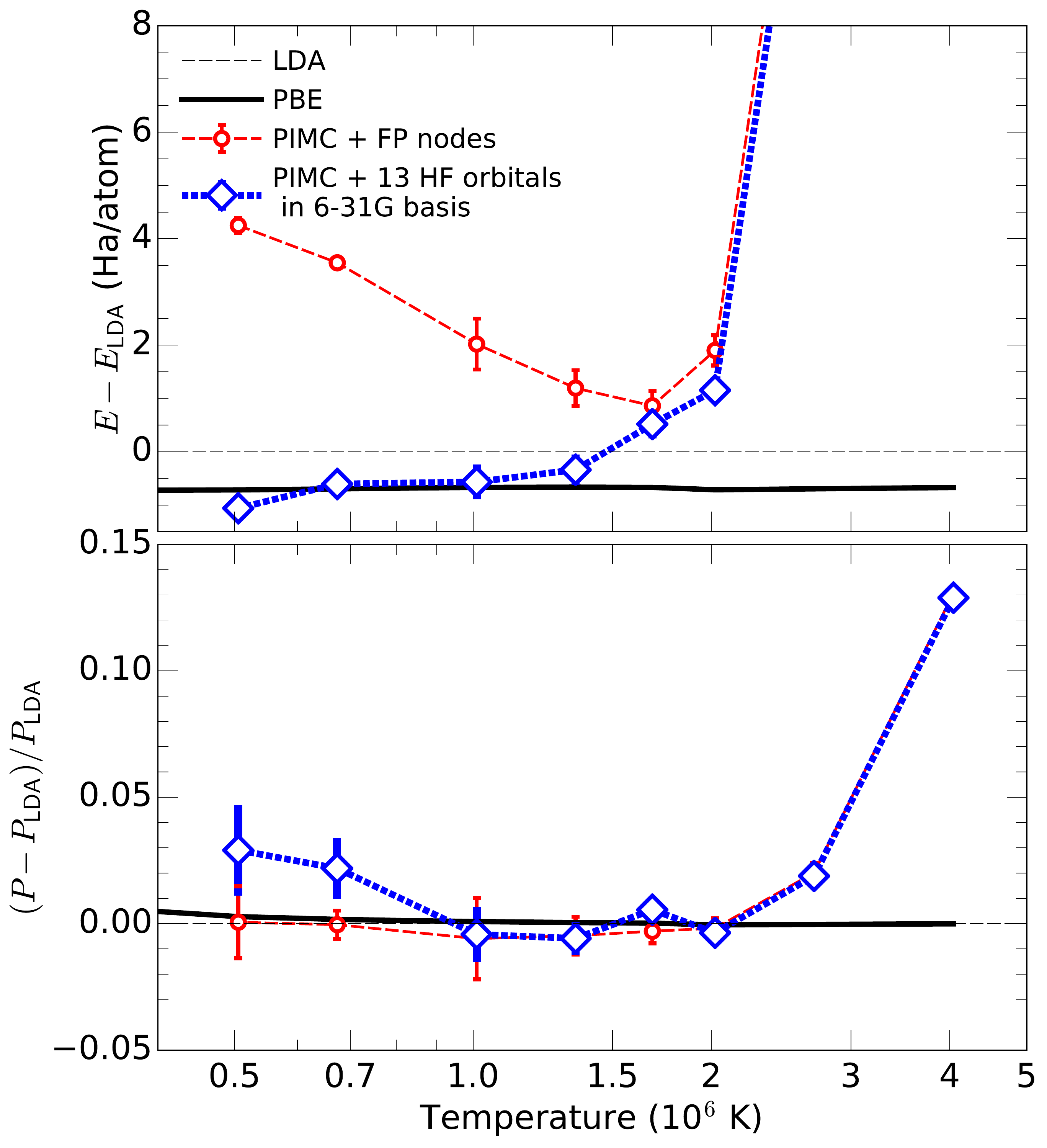}
\caption{\label{fig2} Differences in internal energies and pressures
  of a single Na atom in a periodic 5-Bohr cubic cell calculated using
  different methods relative to those within local density
  approximation (LDA). Dark solid lines represent
  DFT results obtained with the
  Perdew-Burke-Ernzerhof (PBE) functional. 
  Colored dashed lines represent PIMC
  results with free-particle and Hartree-Fock nodes.
  {The LDA and PBE calculations do not include any 
  excitation of the 1s electrons, which explains 
  the deviations from the PIMC results for $T>2\times10^6$ K.}}
\end{figure}

In Ref.~\onlinecite{Militzer2015}, we have shown that accuracy of PIMC simulations can be improved by introducing atomic orbitals, derived with the HF method, to the fermion nodes. 
Here we investigate whether any further improvement
can be made by representing the orbitals with
more accurate basis sets, including a large number
of localized orbitals or by deriving the orbitals with 
LDA or generalized gradient approximation in the 
Perdew-Burke-Ernzerhof (PBE) type \cite{Perdew96}, instead of HF. Before performing many-atom simulations, 
we tested various methods of introducing atomic 
orbitals to study one stationary sodium atom
in a periodic 5-Bohr 
cubic box, and compared the results in Fig. \ref{fig2} 
and Table \ref{table:sodium_atom}.

\begin{table}
\caption{Comparison of energies and pressures calculated using different methods. For the PIMC calculations, we also specify the node type, the basis of localized orbitals and their numbers $n_s$. The internal energies are in units of Ha/atom, pressures are in GPa.}
\label{table:sodium_atom}
\scriptsize
\begin{tabular} {llrcccccc}
\hline
\hline
\
Method & Basis & $n_s$ & $E$~~ & $\sigma_E$~ &  $E$-$E_{\textrm LDA}$ & $P$~~      & $\sigma_P$~ & $P$-$P_{\textrm LDA}$\\
\hline
\multicolumn{2}{l}{$T$=2020958 K} &\\
\hline
LDA        &        &    &  -60.61 &   &     &   11354.6  &     &      \\ 
PBE        &        &    &  -61.32 &   &   -0.71 &   11349.6  &    &    -4.9 \\ 
\hline
PIMC-FP     &        &  0 &  -58.67 & 0.46  &    1.93 &   11356.4  &   68.1 &     1.8 \\ 
\hline
PIMC-HF     & 6-31G  &  6 &  -59.06 & 0.37  &    1.55 &   11357.7  &   56.0 &     3.1 \\ 
PIMC-HF     & 6-31G  & 13 &  -59.45 & 0.19  &    1.16 &   11313.6  &   29.7 &   -41.0 \\ 
PIMC-HF     & 6-31+G & 13 &  -59.15 & 0.27  &    1.46 &   11362.1  &   42.6 &     7.5 \\ 
PIMC-HF     & 6-31+G & 17 &  -59.13 & 0.31  &    1.48 &   11360.8  &   49.3 &     6.2 \\ 
\hline
PIMC-LDA    & 6-31G  & 13 &  -59.77 & 0.43  &    0.83 &   11283.4  &   66.3 &   -71.2 \\ 
PIMC-PBE    & 6-31G  & 13 &  -59.28 & 0.34  &    1.32 &   11340.0  &   53.0 &   -14.6 \\ 
PIMC-PBEX   & 6-31G  & 13 &  -58.92 & 0.39  &    1.69 &   11405.4  &   59.9 &    50.8 \\ 
\hline
\\
\multicolumn{2}{l}{$T$=1010479 K} &\\
\hline
LDA        &        &    & -112.87 & 0.00  &   0.00  &    4528.0 &     0.0 &     0.0 \\ 
PBE        &        &    & -113.54 & 0.00  &  -0.67  &    4531.8 &     0.0 &     3.8 \\ 
\hline
PIMC-FP     &        &  0 & -111.14 & 0.80  &   1.73  &    4474.0 &   125.2 &   -54.0 \\ 
\hline
PIMC-HF     & 6-31G  &  6 & -113.12 & 0.33  &  -0.24  &    4544.7 &    51.5 &    16.7 \\ 
PIMC-HF     & 6-31G  & 13 & -113.44 & 0.29  &  -0.57  &    4508.9 &    45.8 &   -19.1 \\ 
PIMC-HF     & 6-31+G & 13 & -113.47 & 0.23  &  -0.60  &    4508.2 &    37.0 &   -19.8 \\ 
PIMC-HF     & 6-31+G & 17 & -112.73 & 0.31  &   0.15  &    4627.1 &    49.9 &    99.1 \\ 
\hline
PIMC-LDA    & 6-31G  & 13 & -113.53 & 0.36  &  -0.65  &    4473.6 &    55.7 &   -54.4 \\ 
PIMC-PBE    & 6-31G  & 13 & -113.67 & 0.32  &  -0.80  &    4466.6 &    52.6 &   -61.4 \\ 
PIMC-PBEX   & 6-31G  & 13 & -113.38 & 0.32  &  -0.51  &    4507.5 &    50.8 &   -20.5 \\ 
\hline
\hline
\end{tabular}
\end{table}

Figure~\ref{fig2} shows the difference in internal energy and pressure
between PIMC and DFT-MD calculations.  At temperatures of $T>2
\times 10^6$ K, all PIMC energies and pressures are systematically
higher than our DFT-MD results within LDA or PBE. This can be
attributed to the excitations of 1s electrons, which are
treated explicitly as the other outer-shell electrons in PIMC, 
but are frozen in the pseudopotential of DFT-MD calculations.

At temperatures of $T \le 2 \times 10^6$ K, the energies
from PIMC computations with FP nodes are systematically too high,
while the pressures remain quite reasonable. 
In contrast, using PIMC with HF nodes, 
the results are in much better agreement with PBE
predictions.
This suggests that FP nodes do not lead to the 
correct $K$ and $L$ shell structures
of the atom, which is primarily a local property, so
the error in the energy does not depend significantly on density.
{At $T<10^6$ K, pressures from our PIMC+13 HF simulations 
are high, so are the uncertainties. This is due to the low 
sampling efficiency of PIMC at these temperatures.}

Table \ref{table:sodium_atom} lists additional PIMC results for two
temperatures of 1 and 2 million Kelvin.  In all cases, the orbitals
were derived with spin-restricted {\footnotesize GAMESS} \cite{gamess}
calculations of the Na$^{+}$ ion so that we can use the same orbitals
for both spin channels in PIMC calculations. This is a reasonable
approximation because the spin state is of minor importance 
at high temperatures. This is confirmed by
DFT-MD calculations: within LDA and PBE,
the spin-polarized (5+6) and spin-unpolarized calculations yield
similar results for the temperature range under consideration.

In our PIMC calculations with localized nodal surfaces, we need a
minimum of 6 atomic orbitals to provide at least one bound state for
every electron. Table \ref{table:sodium_atom} shows that we found no
statistically significant difference between using 6 and 13 HF
orbitals in PIMC calculations, for both temperatures.  At
1010479 K, the PIMC pressure and energy become too high when we
increase the number of orbitals to 17.  We attribute this deviation to
our small simulation cell of 5.0 Bohr. The highest atomic
orbitals are too delocalized to fit into this cell.

We find no statistically
significant differences in the PIMC results between using a 6-31G and
a slightly more accurate 6-31+G basis set. Both yield similar HF
energies that are 0.176 Ha higher than basis set-converged HF
calculations with a TZV basis. This difference is within the 
error bar of typical PIMC calculations.

We also test whether there would be an
advantage in deriving the atomic orbitals with LDA or PBE methods
rather than with HF theory, and find no significant
difference. Furthermore, our studies with orbitals that are derived
with just PBE exchange (PBEX) lead to similar PIMC results.

In the following many-atom PIMC calculations,
we choose 13 HF orbitals that are generated with 6-31G basis in
Eq. \ref{pollock}.

\subsection{Equation of state}

\begin{figure}
\centering\includegraphics[width=0.5\textwidth]{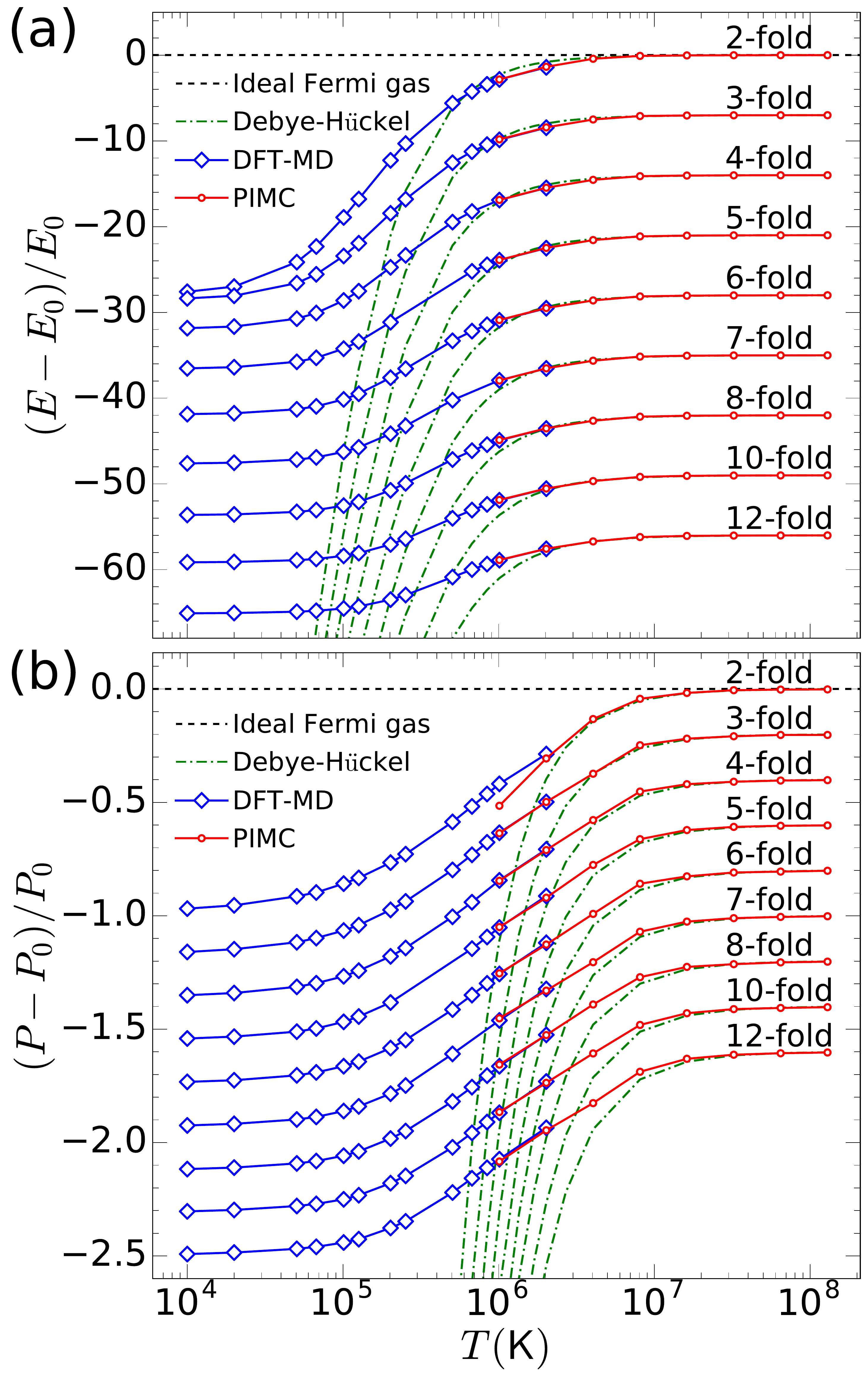}
\caption{\label{fig3} Excess internal energies and pressures, relative
  to the ideal Fermi gas, for a 8-atom simulations
  along nine isochores (2, $\cdots$, 12-fold compression of ambient density). 
  Corresponding
  results of the Debye-H\"{u}ckel theory are plotted for
  comparison.  For clarity, each energy curve 
  at three-times-ambient or higher density is shifted by -7, and
  the pressure curve is shifted by -0.2, with respect to the
  nearby one with lower density. PIMC and DFT-MD predict consistent results
  at 2$\times$10$^6$ K and PIMC agrees with analytical models at above
  2$\times$10$^7$ K.}
\end{figure}

Figure \ref{fig3} shows the calculated internal energies and pressures 
along nine isochores relative to the ideal Fermi electron gas model
\footnote{Here, we include the kinetic energy of the nuclei when
  calculating the total internal energy of the ideal Fermi electron
  gas model, but have not considered the charge and interaction of the
  nuclei.}.  Our results show excellent agreement between PIMC and
DFT-MD at 2$\times$10$^6$ K for all densities. The difference between
PIMC and DFT-MD is typically less than $3$ Ha/atom in internal energy
and within 3\% in pressure.  We have thus succeeded in constructing a
coherent first-principles EOS table for warm dense sodium, over a wide
range of densities (2-12 $\rho_\text{ambient}$) and temperatures
($10^3$-$1.29\times10^8$ K).

At high temperatures of $T>2\times10^7$ K, our PIMC results agree with
those from the Debye-H\"{u}ckel model as well as the Fermi electron
gas theory, because the temperatures are so high that the atoms are
fully ionized.

PIMC successfully bridges DFT-MD at
2$\times$10$^6$ K and analytical models in the high-temperature limit.
This cross-validates the EOS data from present PIMC simulations with
fixed-node approximation, and the use of a frozen core and a
zero-temperature exchange-correlation functional in DFT-MD up 
to $2\times 10^6$ K.

As temperature decreases, the ideal Fermi electron gas model
significantly overestimates the energy and the pressure
because it neglects all interactions.
This is
partially improved in the Debye-H\"{u}ckel model, which 
treats weak interactions correctly within the screening approximation.
However, because the Debye-H\"{u}ckel model still does
not treat bound states, it leads to unphysically low pressures and
energies at low temperatures, as the screening approximation breaks
down and electrons occupy bound states.

\begin{figure}
\centering\includegraphics[width=0.4\textwidth]{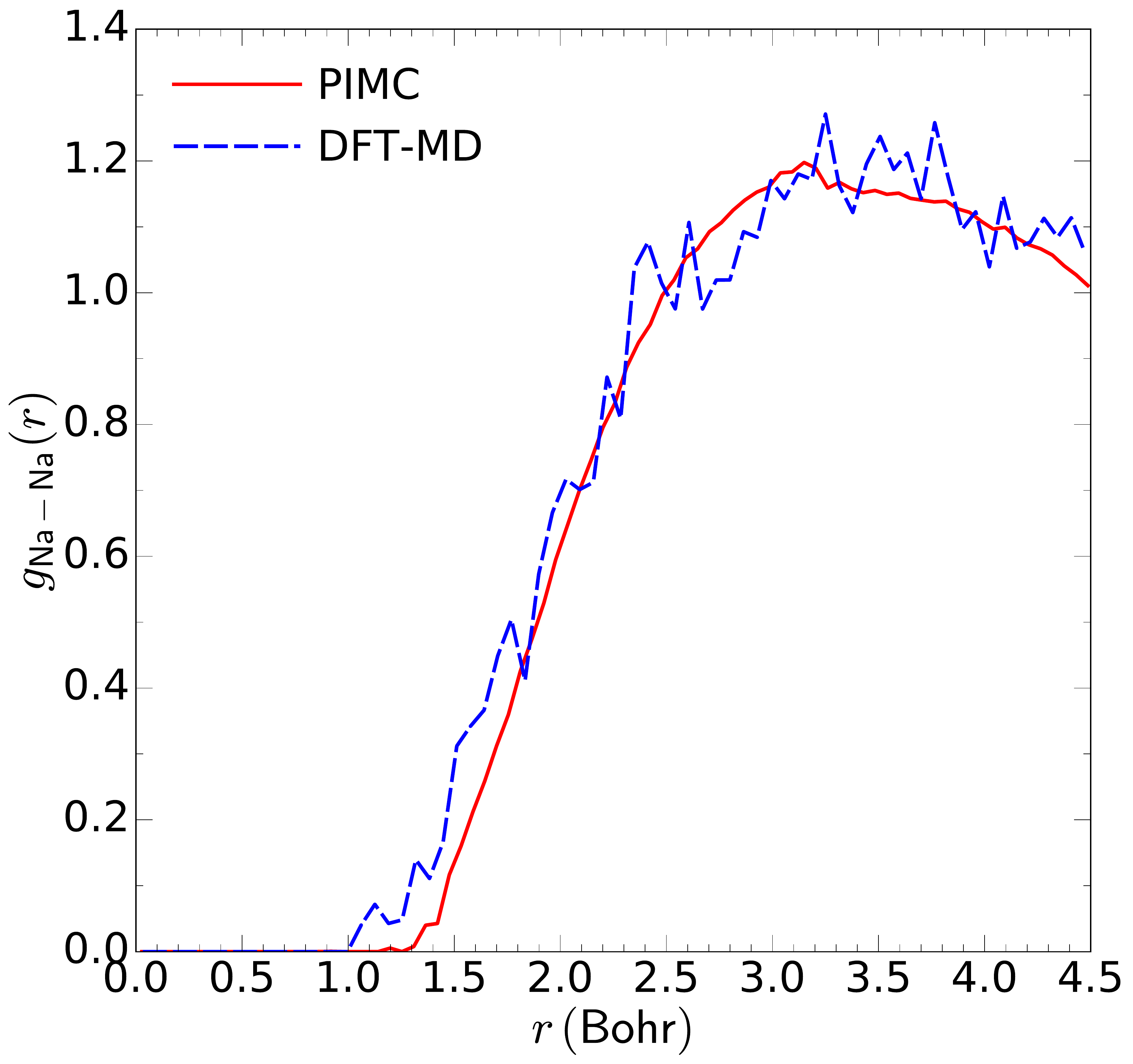}
\caption{\label{fig_gr} Comparison of the nuclear pair correlation
  function obtained from DFT-MD and PIMC for 8-atom simulation cells
  at $10^6$ K and under 8-fold compression.}
\end{figure}

In addition, Fig.~\ref{fig_gr} shows the nuclear pair-correlation functions
$g_\text{Na-Na}(r)$ computed in DFT-MD and PIMC simulations using an
8-atom simulation cell at $10^6$ K and 8-fold compression.  The good
agreement in the $g_\text{Na-Na}(r)$ between the two methods shows
that PIMC and DFT-MD predict consistent ionic plasma structures.  This
is further confirmation that the fixed-node approximation in PIMC and
the exchange-correlation and pseudopotential approximation in DFT-MD
do not inhibit the accuracy of these methods in the WDM regime.

\subsection{Shock compression}
In dynamic compression experiments, the
conservation of mass, momentum, and energy constrains
a steady shock to follow the
Rankine-Hugoniot equation \cite{Ahrens2003}
\begin{equation}\label{eqhug}
\mathcal{H}(T,\rho)=(E-E_0) + \frac{1}{2} (P+P_0)(V-V_0) = 0,
\end{equation}
where ($E_0,P_0,V_0$) and ($E,P,V$) represents the internal energy, pressure, and
volume of the initial and the shocked state, respectively. 
The EOS ($E$ and $P$ on a grid of $T$ and $V$) data 
in the previous section
allows us to solve Eq.~\ref{eqhug} with spline fitting.

Given that samples in shock experiments may be pre-compressed to reach
higher-density, lower-temperature states off the principal Hugoniot,
we consider four different initial conditions 
corresponding to 0.75, 1.0, 1.5, and 2.0 times
$\rho_\textrm{ambient}$.
DFT simulations are performed at each of these densities to
determined the corresponding initial pressures and internal energies.

\begin{figure}
\centering\includegraphics[width=0.5\textwidth]{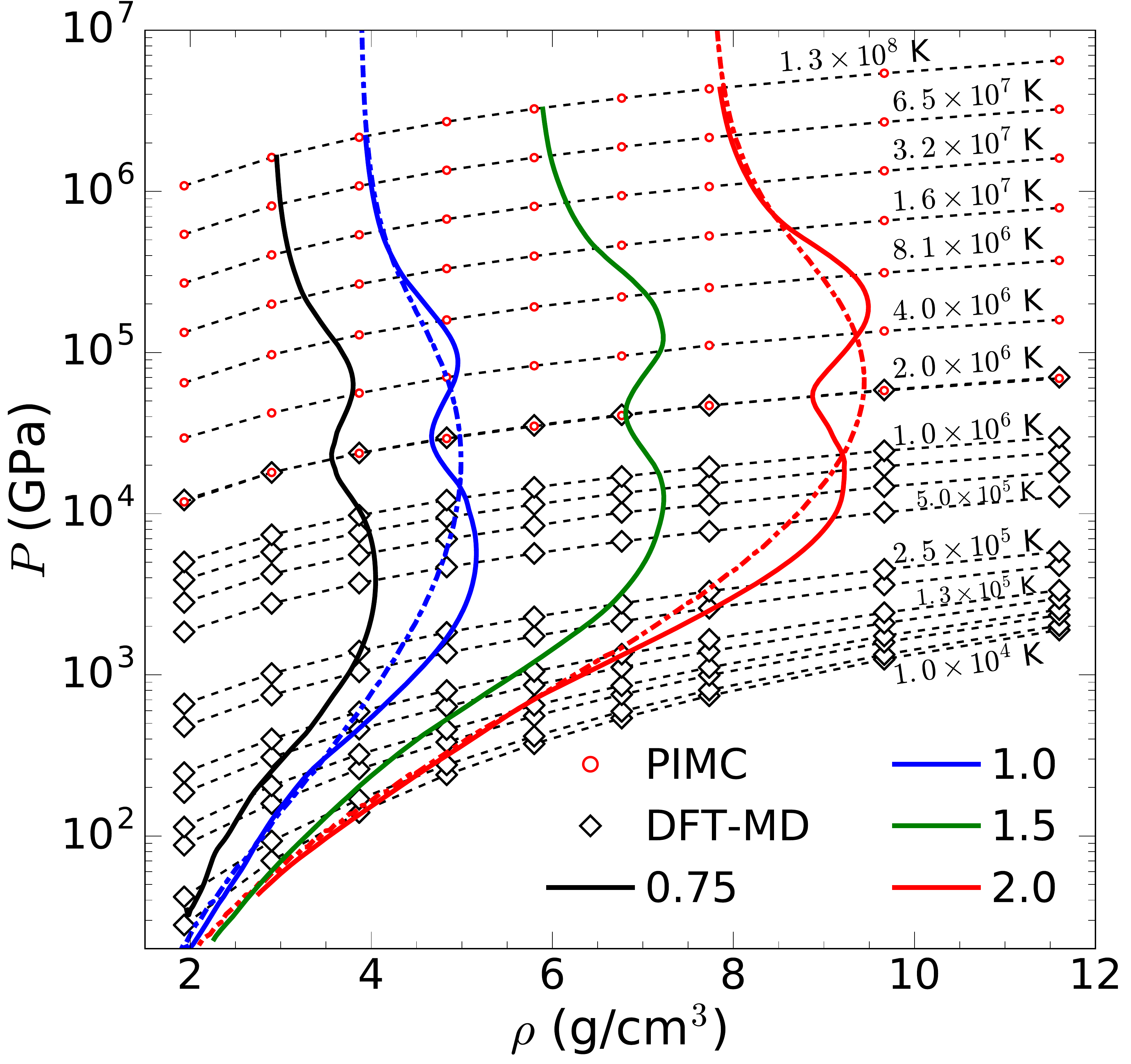}
\caption{\label{fig5} Shock Hugoniot curves (solid curves)
  corresponding to four different initial densities 0.75, 1.0, 1.5,
  and 2 times ambient density represented in
  a pressure-density plot. 
  Corresponding results from SESAME (dashed curves) 
  at two initial densities are plotted for comparison.  
  Isotherms are shown as black dashed lines.}
\end{figure}

We thus calculate the Hugoniot curves and represent them with
pressure-density plots in Fig. \ref{fig5}. 
Two compression peak maxima are predicted along each of
the Hugoniot curves, one above $2\times10^6$ K and the other below $10^6$ K.
We tested different bivariate
interpolation methods for the EOS grid in $\rho-T$ space and 
observed consistent compression curves.
Small discrepancies ($<$0.1) in the compression minimum
in between the two peaks is observed.
We attribute this to the non-smoothness 
caused by the small differences 
between EOS values obtained by PIMC and DFT-MD 
at the corresponding
pressure and temperature conditions.

For comparison, we consider two shock compression
profiles predicted by SESAME \cite{sesame}, which is a tabular
database for the thermodynamic properties of materials that is
constructed by connecting available shock wave data with
Thomas-Fermi-Dirac and Mie-Gr\"{u}neisen theory at high
densities, and some simple analytic forms at low densities. 
The high-temperature
limit is sufficiently described by the Thomas-Fermi-Dirac theory, as
is shown by the remarkable consistency between the SESAME database and
our present first-principles calculations at $T>2\times10^7$
K. However, at lower temperatures when the system includes
bound states, Fermi-Dirac breaks down and the SESAME
database becomes insufficient. SESAME therefore fails to capture shell
effects and only exhibits a single density peak along each Hugoniot curve.

\begin{figure}
\centering\includegraphics[width=0.5\textwidth]{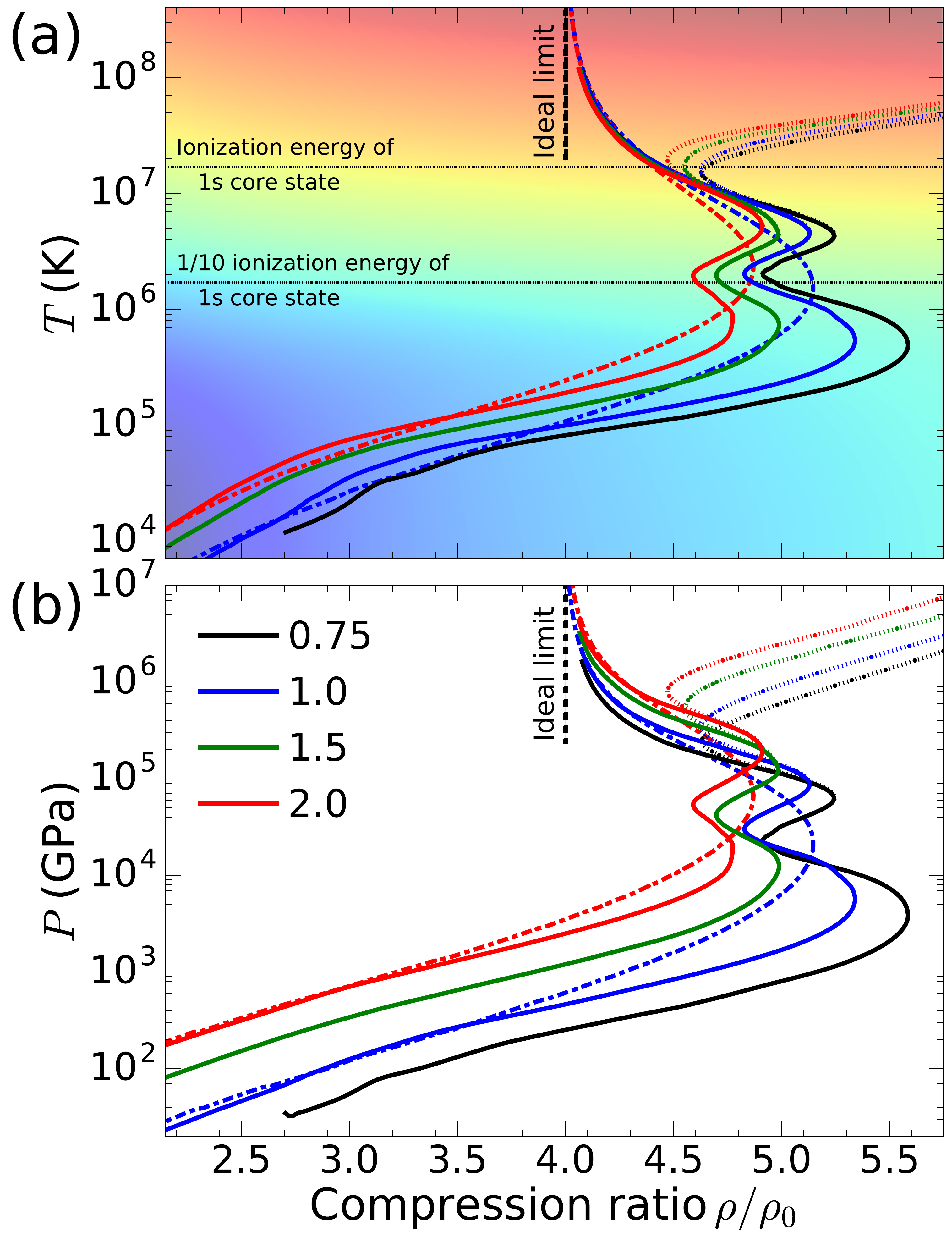}
\caption{\label{fig6} Temperature and pressure as functions of the
  compression ratio along shock Hugoniot curves (solid curves)
  and those with radiation correction (dotted curves)
  corresponding to four initial densities 0.75, 1.0, 1.5, and 2 times
  $\rho_\text{ambient}$.
  Corresponding results from SESAME (dashed curves) 
  at two initial densities are shown for comparison.
  The background color in the upper panel
  reflects the pressure changes.}
\end{figure}

Additionally, Fig. \ref{fig6} shows temperature and pressure along the
Hugoniot curve as functions of the shock compression ratio
$\rho/\rho_0$.  When the initial density is the lowest 
($\rho_0=0.75~\rho_\text{ambient}$), 
a maximum compression ratio of $\sim$5.6 is
reached at 6$\times$10$^5$ K.  The value of this peak decreases with
increasing $\rho_0$, and reduces to 4.7 when $\rho_0=2.0
\rho_\text{ambient}$. The higher-temperature compression peak reaches
slightly larger compression ratio (4.9) than the lower peak as the
initial density increases. In the high-temperature limit, the system is
almost fully ionized and approaches the the non-relativistic ideal
limit ($\rho/\rho_0=4.0$), regardless of the initial density. 

\begin{figure}
\centering\includegraphics[width=0.5\textwidth]{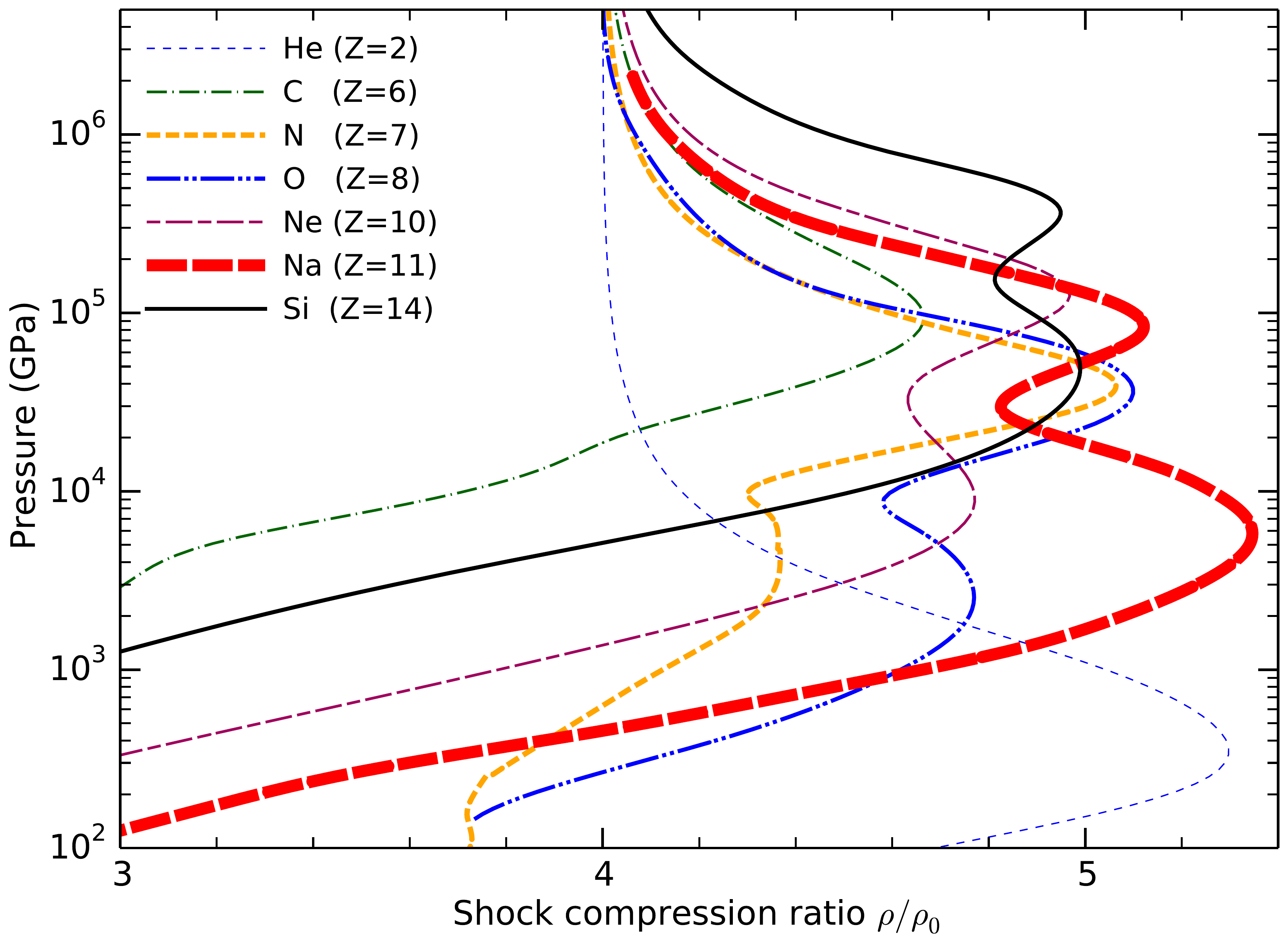}
\caption{\label{fig7} The principal Hugoniot curve of sodium and other elements.}
\end{figure}

Figure \ref{fig7} compares the principal Hugoniot curve of several elements
including He \cite{Militzer2009}, C \cite{Driver2012}, N
\cite{Driver2016Nitrogen}, O \cite{Driver2015Oxygen}, Ne
\cite{Driver2015Neon}, and Si \cite{Militzer2015}.  Interestingly,
sodium together with helium show higher maximum compression ratio than
other elements. We attribute this to the unusual low ambient density
of sodium. In addition, each element imprints a particular structural
signature in its principal Hugoniot curve, exhibiting compression
maxima with different values and at varied pressures and
temperatures, due to the interplay of excitation of internal degrees
of freedom, electronic interactions, and the interaction effects tied
to the initial conditions \cite{PhysRevLett.97.175501}.

Figure \ref{fig0} compares the principal Hugoniot curve of sodium from our
first-principles calculations and those predicted by widely-used EOS
models, SESAME \cite{sesame}, LEOS \cite{leos1,leos2}, and
average-atom Purgatorio (Lynx) \cite{Purgatorio2006,WhitleyLLNL}.  The
compression maxima along the Hugoniot curves are closely related to
ionization and interaction effects. SESAME and LEOS do not explicitly
include information about electronic shell structure, and therefore
do not show two distinct compression maxima.
On the other hand, the
DFT-based, average-atom Purgatorio (Lynx) model does compute the shell
structure for an average of multiple ionization states. Therefore,
Purgatorio (Lynx) agrees well with our first-principles results at above 200
GPa or 10$^5$ K.  Below that, it is less
reliable, because average-atom approaches cannot treat 
bonding and many body effects in a dense fluid 
properly. This is demonstrated by the deviation of the
Hugoniot curve by Purgatorio (Lynx) from our calculations, which agrees with
those predicted by SESAME and LEOS database, which were constructed
by extrapolating experimental
values.  

We also compare our principal Hugoniot results with available
experimental~\cite{Rice1965,Bakanova1965,MarshLASL1980} and
theoretical~\cite{Young1984,Belashchenko2013a} data
available at low temperatures and pressures. 
In comparison with the experimental data,
the local pseudopotential theory~\cite{Young1984} underestimates
the pressure by up to 30 GPa. The EAM model~\cite{Belashchenko2013a}
agrees well with experiments up to 110 GPa, above which the model becomes
invalid. Notably,
DFT-MD, LEOS, and SESAME lie 10-20 GPa below the higest-pressure
experimental data, but all methods tend to agree with the experimental
data at the lowest pressures.

\begin{figure}
\centering\includegraphics[width=0.5\textwidth]{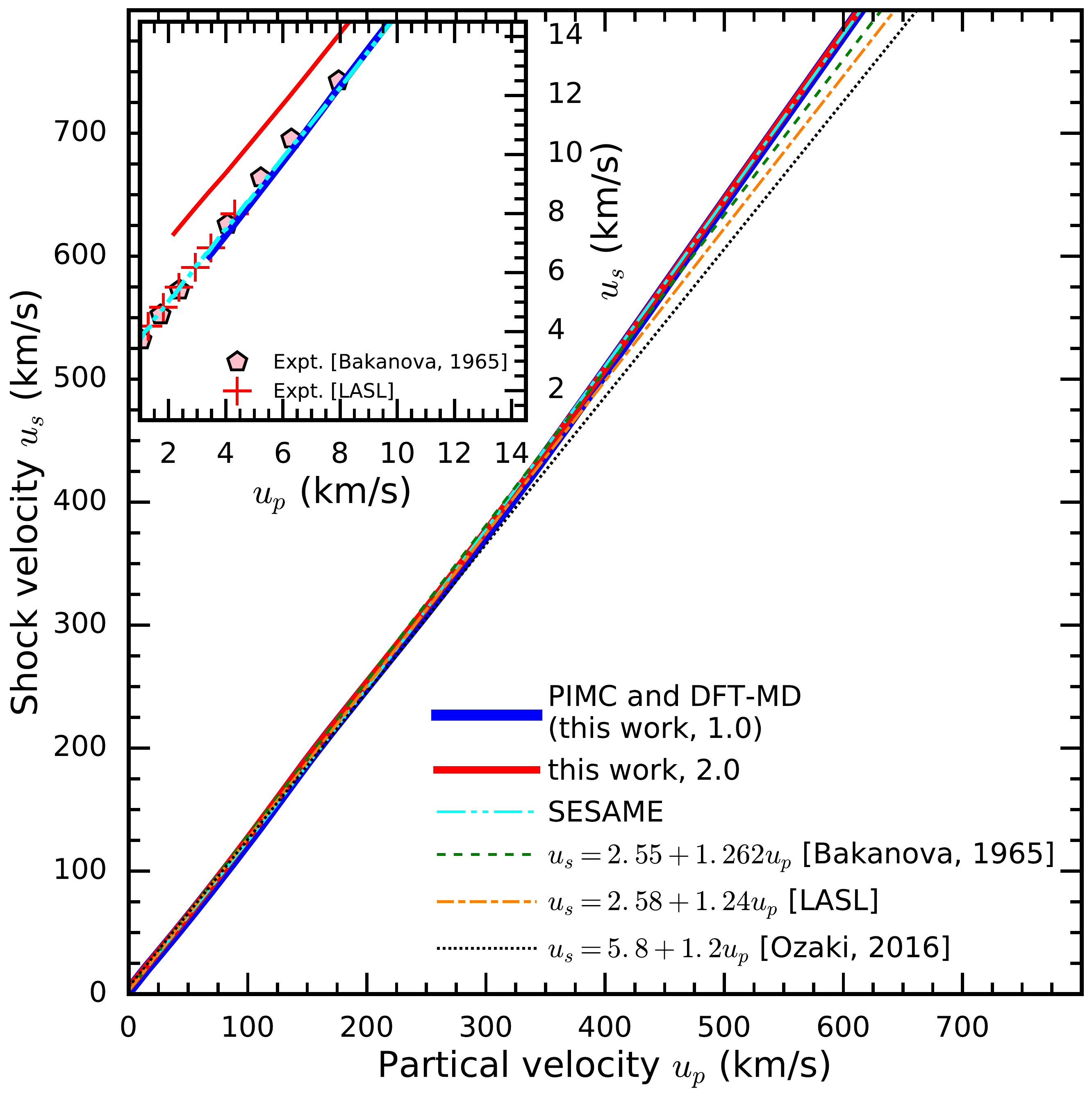}
\caption{\label{usup2} $u_s-u_p$ diagram of sodium
along two Hugoniot curves corresponding to initial
densities of 1.0 and 2.0 times ambient density.
The linear relations found in experiments
are shown in dashed lines for comparison.
The inset is a closer look at 
the low-velocity region.
Shock pressures along the principal Hugoniot curve
are $\sim$1.9 Gbar at shock velocities near 500 km/s.}
\end{figure}

The shock velocity $u_s$ and the particle velocity $u_p$
that are of interests in shock experiments
can be calculated using~\cite{PhysRevLett.87.275502}
\begin{equation}
u_s^2=\xi/\eta ~ \textrm{and} ~ u_p^2=\sqrt{\xi\eta},
\end{equation}
where $\xi=(P-P_0)/\rho_0$, $\eta=1-\rho_0/\rho$. 
Figure~\ref{usup2} compares the relation between $u_s$ and $u_p$
of sodium from our
first-principles calculations and those predicted
by the SESAME database,
low-pressure ($P<10^2$ GPa) shock experiments on sodium~\cite{Bakanova1965,MarshLASL1980},
and an ``universal Hugoniot''~\cite{Ozaki2016}
obtained from high-pressure (up to 0.2 Gbar) experimental
Hugoniot data on Al, Fe, Cu, and Mo.
We find the $u_s-u_p$ relation is
well consistent with the linear relations predicted by
experiments at $u_s<15$ km/s~\cite{Bakanova1965,MarshLASL1980}, 
but increasingly deviates
from them when the shock velocities
exceeds $\sim$500 km/s.
This high velocity corresponds to a shock pressure
of $\sim$1.9 Gbar along the principal Hugoniot curve,
and temperature of $\sim10^7$ K, at which the 
atoms are nearly fully ionized 
(see the discussion in Sec.~\ref{ionization}).
The deviation from the ``universal Hugoniot'' of fluid
metals~\cite{Ozaki2016} is more evident.
This reflects a fundamental difference between the effects
of ionization on shock compression in sodium and 
in heavier metals, such as Fe, Cu, and Mo.
The $u_s-u_p$ profile by SESAME 
agrees remarkably well with our first-principals predictions,
which is not unexpected because of the nearly linear relation 
between $u_s$ and $u_p$ and the consistency between SESAME and
our Hugoniot data at low ($< 10^4$ K) and high ($>10^7$ K) 
temperature and pressure conditions (Figs.~\ref{fig5}-\ref{fig6}). 
Starting from pre-compressed sodium at 2 times ambient
density, the shock velocity is higher but 
the slope $du_s/du_p$ remains 1.31, according to linear
interpolation.

When temperature exceeds 10$^7$ K, the radiation contribution becomes
important. In order to evaluate the radiation effect on shock
compression, we consider an ideal black body scenario and
add the photon contribution to the EOS using
\begin{equation}\label{radiationPE}
P_\textrm{radiation}=\frac{4\sigma}{3c}T^4 ~ \textrm{and} ~ E_\textrm{radiation}=3P_\textrm{radiation}V,
\end{equation}
where $\sigma$ is the Stefan-Boltzmann constant
and $c$ is the speed of light in vacuum.
We then re-construct the Hugoniot curves and show them Figs. \ref{fig0} and \ref{fig6}.
We find that the two compression peaks remain unchanged 
as we include radiative effects. However, the Hugoniot curves 
deviate significantly from the classical limit above $10^7$ K 
and tend towards a compression ratio asymptote of 7.
These results imply that the radiation contribution plays
a significant role in the shock compression of materials at extreme
temperatures ($T>10^7$ K) and pressures ($P>1$ Gbar).  In
comparison, the pressure-contribution from relativistic effects,
included in the Purgatorio (Lynx) model is much smaller (Fig. \ref{fig0}).

\subsection{Ionization}
\label{ionization}



\begin{figure}
\centering\includegraphics[width=0.5\textwidth]{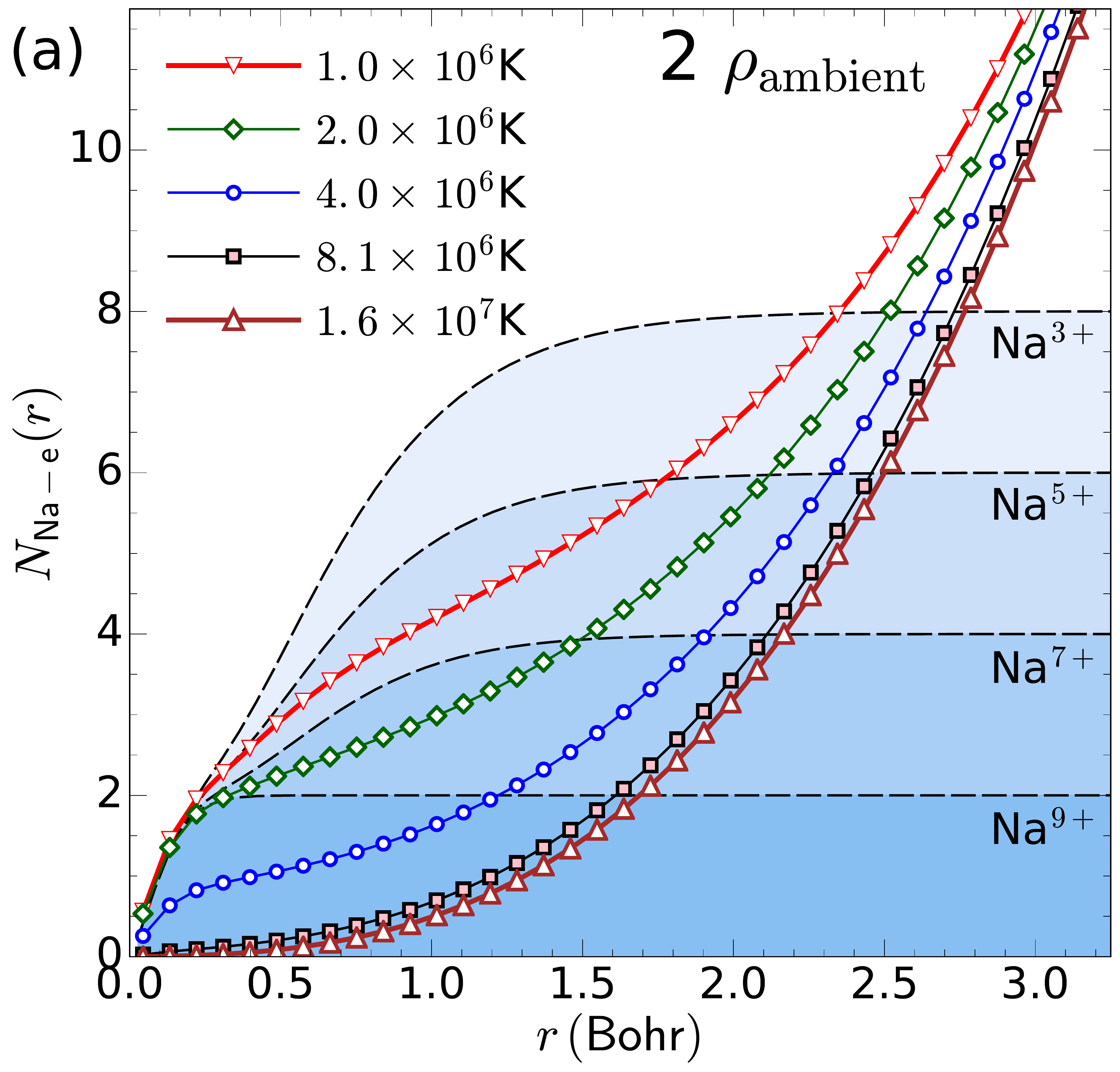}
\centering\includegraphics[width=0.5\textwidth]{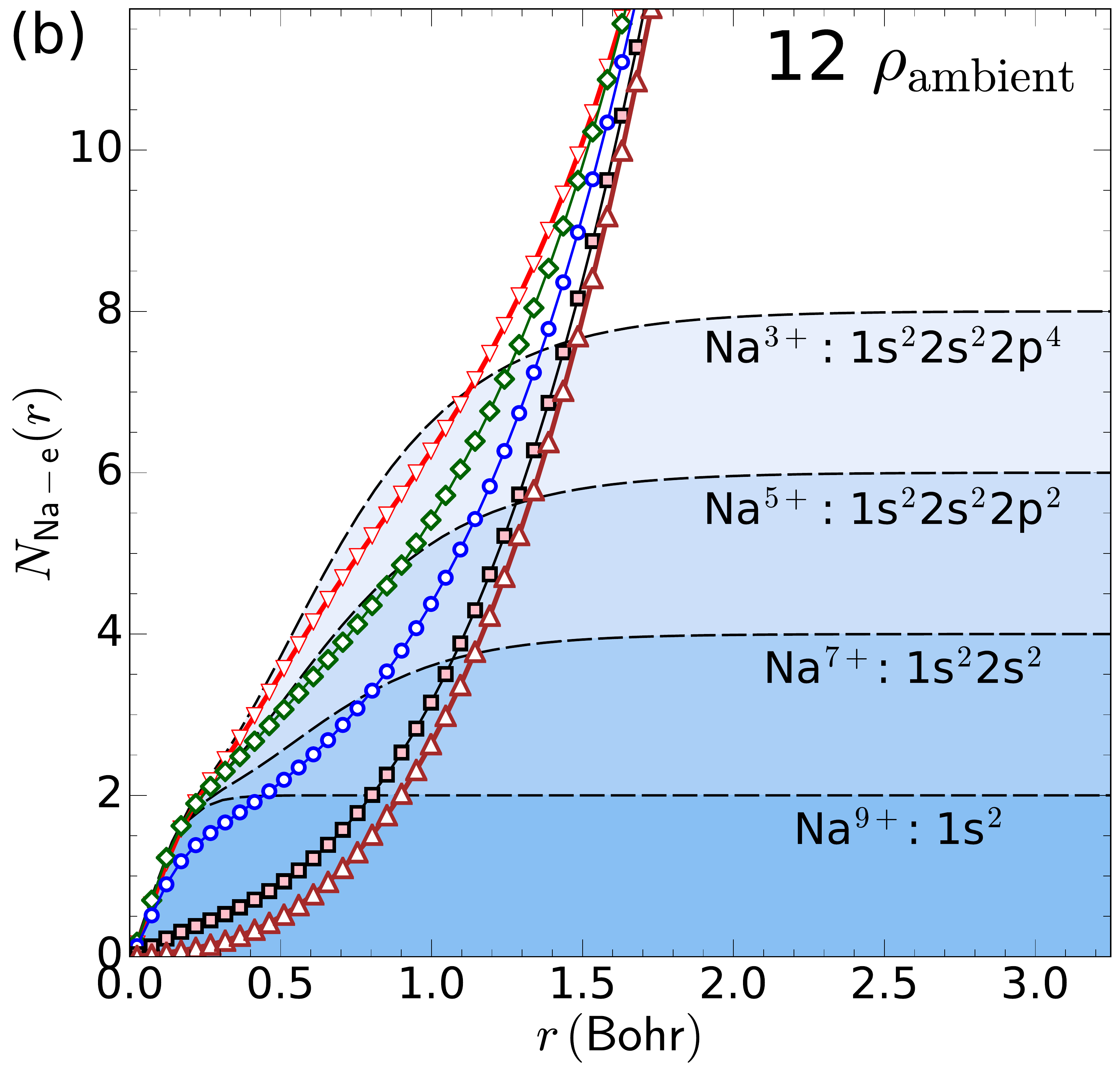}
\caption{\label{fig9}Number of electrons near each nucleus at two
  densities ($\rho=2$ and 12 times $\rho_{\textrm{ambient}}$) and a
  series of increasing temperatures. The profiles of four different
  ionization states of an isolated sodium atom, 
  calculated with {\footnotesize GAMESS}), are
  shown for comparison.}
\end{figure}

The structure of the Hugoniot curves observed in the previous
section can be understood from the
density-temperature dependence of the ionization process.
In order to examine this relation, 
Fig.~\ref{fig9} shows the number of
electrons near the nucleus for a given temperature and density, 
$N_{\textrm{Na}-e}(r)$, given by the formula
\begin{equation}
N(r) = \left< \frac{1}{N_I} \sum_{e,I}
\theta(r-\left|\vec{r}_e-\vec{r}_I \right|) \right>,
\end{equation}
where the sum includes all electron-ion pairs and $\theta$ represents
the Heaviside function. 

The $N_{\textrm{Na}-e}(r)$ functions at two different densities
and a series of temperatures are compared with 
ionization states of isolated sodium ions
(calculated with {\footnotesize GAMESS}).  The results show that
$N_{\textrm{Na}-e}(r)$ decreases with $T$, indicating the gradual
ionization of the atoms with temperature. The energy levels of the 1s
electrons are much deeper than those of the outer-shell electrons, and
therefore only become excited above $2\times10^6$ K.  This indicates
that the upper peaks on the Hugoniot curve in Figs. \ref{fig5} and
\ref{fig6} are related to the excitation of the $K$-shell electrons, while
the lower peaks correspond to the excitation of $L$-shell electrons.


\section{Conclusions}\label{conclusion}
In this work, we construct a thermal density matrix with localized, HF
orbitals to construct fermion nodal surfaces and perform PIMC simulations 
of the second-row element sodium. We obtain an accurate EOS 
from temperatures of 129 million to 1 million K, 
at which the results are consistent with DFT-MD. 
The excellent agreement between the PIMC and DFT-MD 
validates the use of the pseudopotential that freezes 
1s electrons and the use of zero-temperature exchange-correlation 
functionals up to temperateures of $2\times10^6$ K.

By investigating the shock compression curves using the obtained
EOS data, we find two compression maxima along the Hugoniot
curves. This is in contrast to the single-peak Hugoniot curve
predicted by the SESAME and the LEOS database, which are based on
models that neglect bonding and many-body effects.  The higher-temperature
compression maxima occurs at $6\times10^6$ K due to the significant
excitation of 1s electrons, while the lower compression maximum is due
to the thermal excitation of outer-shell electrons. 
We predict a maximum compression ratio of 5.3
along the principal Hugoniot, and compression ratios of greater than 5
can be reached when the initial density is less than
1.5 $\rho_\text{ambient}$. The value decreases at higher initial
densities due to stronger particle interaction, and vice versa.
Including a radiation contribution, shock Hugonits are modified
significantly above 10$^7$~K and 1 Gbar and a much higher compression ratio can
be reached.

\section{Supplementary Material}
See the supplementary material for the tables of EOS data.

\begin{acknowledgments}
  
  This research is supported by the U. S. Department of Energy, grant
  DE-SC0010517 and DE-SC0016248. 
Shuai Zhang is partially supported by the PLS-Postdoctoral
Grant of the Lawrence Livermore National Laboratory.
Computational support was provided by the
  Janus supercomputer, which is supported by the National Science
  Foundation (Grant No. CNS-0821794), the University of Colorado, and
  the National Center for Atmospheric Research, and NERSC. This research 
  is part of the Blue Waters sustained-petascale computing project 
  (NSF ACI 1640776), which 
  is supported by the National Science Foundation 
  (awards OCI-0725070 and ACI-1238993) and the state of Illinois. 
  Blue Waters is a joint effort of the University of Illinois at 
  Urbana-Champaign and its National Center for Supercomputing Applications. S.Z. appreciates the helpful discussion with Mu Li.
  
\end{acknowledgments}



\begin{thebibliography}{95}%
\makeatletter
\providecommand \@ifxundefined [1]{%
 \@ifx{#1\undefined}
}%
\providecommand \@ifnum [1]{%
 \ifnum #1\expandafter \@firstoftwo
 \else \expandafter \@secondoftwo
 \fi
}%
\providecommand \@ifx [1]{%
 \ifx #1\expandafter \@firstoftwo
 \else \expandafter \@secondoftwo
 \fi
}%
\providecommand \natexlab [1]{#1}%
\providecommand \enquote  [1]{``#1''}%
\providecommand \bibnamefont  [1]{#1}%
\providecommand \bibfnamefont [1]{#1}%
\providecommand \citenamefont [1]{#1}%
\providecommand \href@noop [0]{\@secondoftwo}%
\providecommand \href [0]{\begingroup \@sanitize@url \@href}%
\providecommand \@href[1]{\@@startlink{#1}\@@href}%
\providecommand \@@href[1]{\endgroup#1\@@endlink}%
\providecommand \@sanitize@url [0]{\catcode `\\12\catcode `\$12\catcode
  `\&12\catcode `\#12\catcode `\^12\catcode `\_12\catcode `\%12\relax}%
\providecommand \@@startlink[1]{}%
\providecommand \@@endlink[0]{}%
\providecommand \url  [0]{\begingroup\@sanitize@url \@url }%
\providecommand \@url [1]{\endgroup\@href {#1}{\urlprefix }}%
\providecommand \urlprefix  [0]{URL }%
\providecommand \Eprint [0]{\href }%
\providecommand \doibase [0]{http://dx.doi.org/}%
\providecommand \selectlanguage [0]{\@gobble}%
\providecommand \bibinfo  [0]{\@secondoftwo}%
\providecommand \bibfield  [0]{\@secondoftwo}%
\providecommand \translation [1]{[#1]}%
\providecommand \BibitemOpen [0]{}%
\providecommand \bibitemStop [0]{}%
\providecommand \bibitemNoStop [0]{.\EOS\space}%
\providecommand \EOS [0]{\spacefactor3000\relax}%
\providecommand \BibitemShut  [1]{\csname bibitem#1\endcsname}%
\let\auto@bib@innerbib\@empty
\bibitem [{\citenamefont {Wigner}\ and\ \citenamefont
  {Seitz}(1933)}]{Wigner1933}%
  \BibitemOpen
  \bibfield  {author} {\bibinfo {author} {\bibfnamefont {E.~P.}\ \bibnamefont
  {Wigner}}\ and\ \bibinfo {author} {\bibfnamefont {F.}~\bibnamefont {Seitz}},\
  }\href@noop {} {\bibfield  {journal} {\bibinfo  {journal} {Phys. Rev.}\
  }\textbf {\bibinfo {volume} {43}},\ \bibinfo {pages} {804} (\bibinfo {year}
  {1933})}\BibitemShut {NoStop}%
\bibitem [{\citenamefont {Neaton}\ and\ \citenamefont
  {Ashcroft}(2001)}]{Neaton2001}%
  \BibitemOpen
  \bibfield  {author} {\bibinfo {author} {\bibfnamefont {J.}~\bibnamefont
  {Neaton}}\ and\ \bibinfo {author} {\bibfnamefont {N.}~\bibnamefont
  {Ashcroft}},\ }\href@noop {} {\bibfield  {journal} {\bibinfo  {journal}
  {Phys. Rev. Lett.}\ }\textbf {\bibinfo {volume} {86}},\ \bibinfo {pages}
  {2830} (\bibinfo {year} {2001})}\BibitemShut {NoStop}%
\bibitem [{\citenamefont {Eremets}(1996)}]{Eremets1996}%
  \BibitemOpen
  \bibfield  {author} {\bibinfo {author} {\bibfnamefont {M.}~\bibnamefont
  {Eremets}},\ }\href@noop {} {\emph {\bibinfo {title} {High Pressure
  Experimental Methods Oxford Science Publications}}}\ (\bibinfo  {publisher}
  {Oxford University Press, Oxford},\ \bibinfo {year} {1996})\BibitemShut
  {NoStop}%
\bibitem [{\citenamefont {Dubrovinsky}\ \emph {et~al.}(2012)\citenamefont
  {Dubrovinsky}, \citenamefont {Dubrovinskaia}, \citenamefont {Prakapenka},\
  and\ \citenamefont {Abakumov}}]{Dubrovinsky2012}%
  \BibitemOpen
  \bibfield  {author} {\bibinfo {author} {\bibfnamefont {L.}~\bibnamefont
  {Dubrovinsky}}, \bibinfo {author} {\bibfnamefont {N.}~\bibnamefont
  {Dubrovinskaia}}, \bibinfo {author} {\bibfnamefont {V.~B.}\ \bibnamefont
  {Prakapenka}}, \ and\ \bibinfo {author} {\bibfnamefont {A.~M.}\ \bibnamefont
  {Abakumov}},\ }\href@noop {} {\bibfield  {journal} {\bibinfo  {journal}
  {Nature Commun.}\ }\textbf {\bibinfo {volume} {3}},\ \bibinfo {pages} {1163}
  (\bibinfo {year} {2012})}\BibitemShut {NoStop}%
\bibitem [{\citenamefont {Gregoryanz}\ \emph {et~al.}(2008)\citenamefont
  {Gregoryanz}, \citenamefont {Lundegaard}, \citenamefont {McMahon},
  \citenamefont {Guillaume}, \citenamefont {Nelmes},\ and\ \citenamefont
  {Mezouar}}]{Gregoryanz2008}%
  \BibitemOpen
  \bibfield  {author} {\bibinfo {author} {\bibfnamefont {E.}~\bibnamefont
  {Gregoryanz}}, \bibinfo {author} {\bibfnamefont {L.~F.}\ \bibnamefont
  {Lundegaard}}, \bibinfo {author} {\bibfnamefont {M.~I.}\ \bibnamefont
  {McMahon}}, \bibinfo {author} {\bibfnamefont {C.}~\bibnamefont {Guillaume}},
  \bibinfo {author} {\bibfnamefont {R.~J.}\ \bibnamefont {Nelmes}}, \ and\
  \bibinfo {author} {\bibfnamefont {M.}~\bibnamefont {Mezouar}},\ }\href@noop
  {} {\bibfield  {journal} {\bibinfo  {journal} {Science}\ }\textbf {\bibinfo
  {volume} {320}},\ \bibinfo {pages} {1054} (\bibinfo {year}
  {2008})}\BibitemShut {NoStop}%
\bibitem [{\citenamefont {Aleksandrov}\ \emph {et~al.}(1982)\citenamefont
  {Aleksandrov}, \citenamefont {Kachinskii}, \citenamefont {Makarenko},\ and\
  \citenamefont {Stishov}}]{Aleksandrov1982}%
  \BibitemOpen
  \bibfield  {author} {\bibinfo {author} {\bibfnamefont {I.}~\bibnamefont
  {Aleksandrov}}, \bibinfo {author} {\bibfnamefont {V.}~\bibnamefont
  {Kachinskii}}, \bibinfo {author} {\bibfnamefont {I.}~\bibnamefont
  {Makarenko}}, \ and\ \bibinfo {author} {\bibfnamefont {S.}~\bibnamefont
  {Stishov}},\ }\href@noop {} {\bibfield  {journal} {\bibinfo  {journal} {ZhETF
  Pisma Redaktsiiu}\ }\textbf {\bibinfo {volume} {36}},\ \bibinfo {pages} {336}
  (\bibinfo {year} {1982})}\BibitemShut {NoStop}%
\bibitem [{\citenamefont {Fritz}\ and\ \citenamefont
  {Olinger}(1984)}]{Fritz1984}%
  \BibitemOpen
  \bibfield  {author} {\bibinfo {author} {\bibfnamefont {J.~N.}\ \bibnamefont
  {Fritz}}\ and\ \bibinfo {author} {\bibfnamefont {B.}~\bibnamefont
  {Olinger}},\ }\href@noop {} {\bibfield  {journal} {\bibinfo  {journal} {J.
  Chem. Phys.}\ }\textbf {\bibinfo {volume} {80}},\ \bibinfo {pages} {2864}
  (\bibinfo {year} {1984})}\BibitemShut {NoStop}%
\bibitem [{\citenamefont {Hanfland}, \citenamefont {Loa},\ and\ \citenamefont
  {Syassen}(2002)}]{Hanfland2002}%
  \BibitemOpen
  \bibfield  {author} {\bibinfo {author} {\bibfnamefont {M.}~\bibnamefont
  {Hanfland}}, \bibinfo {author} {\bibfnamefont {I.}~\bibnamefont {Loa}}, \
  and\ \bibinfo {author} {\bibfnamefont {K.}~\bibnamefont {Syassen}},\ }\href
  {\doibase 10.1103/PhysRevB.65.184109} {\bibfield  {journal} {\bibinfo
  {journal} {Phys. Rev. B}\ }\textbf {\bibinfo {volume} {65}},\ \bibinfo
  {pages} {184109} (\bibinfo {year} {2002})}\BibitemShut {NoStop}%
\bibitem [{\citenamefont {Raty}, \citenamefont {Schwegler},\ and\ \citenamefont
  {Bonev}(2007)}]{Raty2007}%
  \BibitemOpen
  \bibfield  {author} {\bibinfo {author} {\bibfnamefont {J.-Y.}\ \bibnamefont
  {Raty}}, \bibinfo {author} {\bibfnamefont {E.}~\bibnamefont {Schwegler}}, \
  and\ \bibinfo {author} {\bibfnamefont {S.~A.}\ \bibnamefont {Bonev}},\ }\href
  {\doibase 10.1038/nature06123} {\bibfield  {journal} {\bibinfo  {journal}
  {Nature}\ }\textbf {\bibinfo {volume} {449}},\ \bibinfo {pages} {448}
  (\bibinfo {year} {2007})}\BibitemShut {NoStop}%
\bibitem [{\citenamefont {McMahon}\ \emph {et~al.}(2007)\citenamefont
  {McMahon}, \citenamefont {Gregoryanz}, \citenamefont {Lundegaard},
  \citenamefont {Loa}, \citenamefont {Guillaume}, \citenamefont {Nelmes},
  \citenamefont {Kleppe}, \citenamefont {Amboage}, \citenamefont {Wilhelm},\
  and\ \citenamefont {Jephcoat}}]{McMahon2007}%
  \BibitemOpen
  \bibfield  {author} {\bibinfo {author} {\bibfnamefont {M.~I.}\ \bibnamefont
  {McMahon}}, \bibinfo {author} {\bibfnamefont {E.}~\bibnamefont {Gregoryanz}},
  \bibinfo {author} {\bibfnamefont {L.~F.}\ \bibnamefont {Lundegaard}},
  \bibinfo {author} {\bibfnamefont {I.}~\bibnamefont {Loa}}, \bibinfo {author}
  {\bibfnamefont {C.}~\bibnamefont {Guillaume}}, \bibinfo {author}
  {\bibfnamefont {R.~J.}\ \bibnamefont {Nelmes}}, \bibinfo {author}
  {\bibfnamefont {A.~K.}\ \bibnamefont {Kleppe}}, \bibinfo {author}
  {\bibfnamefont {M.}~\bibnamefont {Amboage}}, \bibinfo {author} {\bibfnamefont
  {H.}~\bibnamefont {Wilhelm}}, \ and\ \bibinfo {author} {\bibfnamefont
  {A.~P.}\ \bibnamefont {Jephcoat}},\ }\href {\doibase 10.1073/pnas.0709309104}
  {\bibfield  {journal} {\bibinfo  {journal} {Proc. Natl. Acad. Sci. USA}\
  }\textbf {\bibinfo {volume} {104}},\ \bibinfo {pages} {17297} (\bibinfo
  {year} {2007})}\BibitemShut {NoStop}%
\bibitem [{\citenamefont {Rousseau}\ \emph {et~al.}(2011)\citenamefont
  {Rousseau}, \citenamefont {Xie}, \citenamefont {Ma},\ and\ \citenamefont
  {Bergara}}]{Rousseau2011}%
  \BibitemOpen
  \bibfield  {author} {\bibinfo {author} {\bibfnamefont {B.}~\bibnamefont
  {Rousseau}}, \bibinfo {author} {\bibfnamefont {Y.}~\bibnamefont {Xie}},
  \bibinfo {author} {\bibfnamefont {Y.}~\bibnamefont {Ma}}, \ and\ \bibinfo
  {author} {\bibfnamefont {A.}~\bibnamefont {Bergara}},\ }\href {\doibase
  10.1140/epjb/e2011-10972-9} {\bibfield  {journal} {\bibinfo  {journal} {Eur.
  Phys. J. B}\ }\textbf {\bibinfo {volume} {81}},\ \bibinfo {pages} {1}
  (\bibinfo {year} {2011})}\BibitemShut {NoStop}%
\bibitem [{\citenamefont {Ma}\ \emph {et~al.}(2009)\citenamefont {Ma},
  \citenamefont {Eremets}, \citenamefont {Oganov}, \citenamefont {Xie},
  \citenamefont {Trojan}, \citenamefont {Medvedev}, \citenamefont {Lyakhov},
  \citenamefont {Valle},\ and\ \citenamefont {Prakapenka}}]{Ma2009}%
  \BibitemOpen
  \bibfield  {author} {\bibinfo {author} {\bibfnamefont {Y.}~\bibnamefont
  {Ma}}, \bibinfo {author} {\bibfnamefont {M.}~\bibnamefont {Eremets}},
  \bibinfo {author} {\bibfnamefont {A.~R.}\ \bibnamefont {Oganov}}, \bibinfo
  {author} {\bibfnamefont {Y.}~\bibnamefont {Xie}}, \bibinfo {author}
  {\bibfnamefont {I.}~\bibnamefont {Trojan}}, \bibinfo {author} {\bibfnamefont
  {S.}~\bibnamefont {Medvedev}}, \bibinfo {author} {\bibfnamefont {A.~O.}\
  \bibnamefont {Lyakhov}}, \bibinfo {author} {\bibfnamefont {M.}~\bibnamefont
  {Valle}}, \ and\ \bibinfo {author} {\bibfnamefont {V.}~\bibnamefont
  {Prakapenka}},\ }\href {\doibase 10.1038/nature07786} {\bibfield  {journal}
  {\bibinfo  {journal} {Nature}\ }\textbf {\bibinfo {volume} {457}},\ \bibinfo
  {pages} {182} (\bibinfo {year} {2009})}\BibitemShut {NoStop}%
\bibitem [{\citenamefont {Lazicki}\ \emph {et~al.}(2009)\citenamefont
  {Lazicki}, \citenamefont {Goncharov}, \citenamefont {Struzhkin},
  \citenamefont {Cohen}, \citenamefont {Liu}, \citenamefont {Gregoryanz},
  \citenamefont {Guillaume}, \citenamefont {Mao},\ and\ \citenamefont
  {Hemley}}]{Lazicki2009}%
  \BibitemOpen
  \bibfield  {author} {\bibinfo {author} {\bibfnamefont {A.}~\bibnamefont
  {Lazicki}}, \bibinfo {author} {\bibfnamefont {A.~F.}\ \bibnamefont
  {Goncharov}}, \bibinfo {author} {\bibfnamefont {V.~V.}\ \bibnamefont
  {Struzhkin}}, \bibinfo {author} {\bibfnamefont {R.~E.}\ \bibnamefont
  {Cohen}}, \bibinfo {author} {\bibfnamefont {Z.}~\bibnamefont {Liu}}, \bibinfo
  {author} {\bibfnamefont {E.}~\bibnamefont {Gregoryanz}}, \bibinfo {author}
  {\bibfnamefont {C.}~\bibnamefont {Guillaume}}, \bibinfo {author}
  {\bibfnamefont {H.-K.}\ \bibnamefont {Mao}}, \ and\ \bibinfo {author}
  {\bibfnamefont {R.~J.}\ \bibnamefont {Hemley}},\ }\href@noop {} {\bibfield
  {journal} {\bibinfo  {journal} {Proc. Natl. Acad. Sci. USA}\ }\textbf
  {\bibinfo {volume} {106}},\ \bibinfo {pages} {6525} (\bibinfo {year}
  {2009})}\BibitemShut {NoStop}%
\bibitem [{\citenamefont {Gatti}, \citenamefont {Tokatly},\ and\ \citenamefont
  {Rubio}(2010)}]{Gatti2010}%
  \BibitemOpen
  \bibfield  {author} {\bibinfo {author} {\bibfnamefont {M.}~\bibnamefont
  {Gatti}}, \bibinfo {author} {\bibfnamefont {I.~V.}\ \bibnamefont {Tokatly}},
  \ and\ \bibinfo {author} {\bibfnamefont {A.}~\bibnamefont {Rubio}},\ }\href
  {\doibase 10.1103/PhysRevLett.104.216404} {\bibfield  {journal} {\bibinfo
  {journal} {Phys. Rev. Lett.}\ }\textbf {\bibinfo {volume} {104}},\ \bibinfo
  {pages} {216404} (\bibinfo {year} {2010})}\BibitemShut {NoStop}%
\bibitem [{\citenamefont {Mao}\ \emph {et~al.}(2011)\citenamefont {Mao},
  \citenamefont {Ding}, \citenamefont {Xiao}, \citenamefont {Chow},
  \citenamefont {Shu}, \citenamefont {Lebègue}, \citenamefont {Lazicki},\ and\
  \citenamefont {Ahuja}}]{Mao20122011}%
  \BibitemOpen
  \bibfield  {author} {\bibinfo {author} {\bibfnamefont {H.-K.}\ \bibnamefont
  {Mao}}, \bibinfo {author} {\bibfnamefont {Y.}~\bibnamefont {Ding}}, \bibinfo
  {author} {\bibfnamefont {Y.}~\bibnamefont {Xiao}}, \bibinfo {author}
  {\bibfnamefont {P.}~\bibnamefont {Chow}}, \bibinfo {author} {\bibfnamefont
  {J.}~\bibnamefont {Shu}}, \bibinfo {author} {\bibfnamefont {S.}~\bibnamefont
  {Lebègue}}, \bibinfo {author} {\bibfnamefont {A.}~\bibnamefont {Lazicki}}, \
  and\ \bibinfo {author} {\bibfnamefont {R.}~\bibnamefont {Ahuja}},\ }\href
  {\doibase 10.1073/pnas.1116930108} {\bibfield  {journal} {\bibinfo  {journal}
  {Proc. Natl. Acad. Sci. USA}\ }\textbf {\bibinfo {volume} {108}},\ \bibinfo
  {pages} {20434} (\bibinfo {year} {2011})}\BibitemShut {NoStop}%
\bibitem [{\citenamefont {Marqu{\'{e}}s}\ \emph {et~al.}(2011)\citenamefont
  {Marqu{\'{e}}s}, \citenamefont {Santoro}, \citenamefont {Guillaume},
  \citenamefont {Gorelli}, \citenamefont {Contreras-Garc{\'{i}}a},
  \citenamefont {Howie}, \citenamefont {Goncharov},\ and\ \citenamefont
  {Gregoryanz}}]{Marques2011}%
  \BibitemOpen
  \bibfield  {author} {\bibinfo {author} {\bibfnamefont {M.}~\bibnamefont
  {Marqu{\'{e}}s}}, \bibinfo {author} {\bibfnamefont {M.}~\bibnamefont
  {Santoro}}, \bibinfo {author} {\bibfnamefont {C.~L.}\ \bibnamefont
  {Guillaume}}, \bibinfo {author} {\bibfnamefont {F.~A.}\ \bibnamefont
  {Gorelli}}, \bibinfo {author} {\bibfnamefont {J.}~\bibnamefont
  {Contreras-Garc{\'{i}}a}}, \bibinfo {author} {\bibfnamefont {R.~T.}\
  \bibnamefont {Howie}}, \bibinfo {author} {\bibfnamefont {A.~F.}\ \bibnamefont
  {Goncharov}}, \ and\ \bibinfo {author} {\bibfnamefont {E.}~\bibnamefont
  {Gregoryanz}},\ }\href@noop {} {\bibfield  {journal} {\bibinfo  {journal}
  {Phys. Rev. B}\ }\textbf {\bibinfo {volume} {83}},\ \bibinfo {pages} {184106}
  (\bibinfo {year} {2011})}\BibitemShut {NoStop}%
\bibitem [{\citenamefont {Loa}\ \emph {et~al.}(2011)\citenamefont {Loa},
  \citenamefont {Syassen}, \citenamefont {Monaco}, \citenamefont {Vank\'o},
  \citenamefont {Krisch},\ and\ \citenamefont {Hanfland}}]{Loa2011}%
  \BibitemOpen
  \bibfield  {author} {\bibinfo {author} {\bibfnamefont {I.}~\bibnamefont
  {Loa}}, \bibinfo {author} {\bibfnamefont {K.}~\bibnamefont {Syassen}},
  \bibinfo {author} {\bibfnamefont {G.}~\bibnamefont {Monaco}}, \bibinfo
  {author} {\bibfnamefont {G.}~\bibnamefont {Vank\'o}}, \bibinfo {author}
  {\bibfnamefont {M.}~\bibnamefont {Krisch}}, \ and\ \bibinfo {author}
  {\bibfnamefont {M.}~\bibnamefont {Hanfland}},\ }\href {\doibase
  10.1103/PhysRevLett.107.086402} {\bibfield  {journal} {\bibinfo  {journal}
  {Phys. Rev. Lett.}\ }\textbf {\bibinfo {volume} {107}},\ \bibinfo {pages}
  {086402} (\bibinfo {year} {2011})}\BibitemShut {NoStop}%
\bibitem [{\citenamefont {Pozzo}, \citenamefont {Desjarlais},\ and\
  \citenamefont {Alf\`e}(2011)}]{Pozzo2011}%
  \BibitemOpen
  \bibfield  {author} {\bibinfo {author} {\bibfnamefont {M.}~\bibnamefont
  {Pozzo}}, \bibinfo {author} {\bibfnamefont {M.~P.}\ \bibnamefont
  {Desjarlais}}, \ and\ \bibinfo {author} {\bibfnamefont {D.}~\bibnamefont
  {Alf\`e}},\ }\href {\doibase 10.1103/PhysRevB.84.054203} {\bibfield
  {journal} {\bibinfo  {journal} {Phys. Rev. B}\ }\textbf {\bibinfo {volume}
  {84}},\ \bibinfo {pages} {054203} (\bibinfo {year} {2011})}\BibitemShut
  {NoStop}%
\bibitem [{\citenamefont {Iba\~{n}ez Azpiroz}\ \emph
  {et~al.}(2014)\citenamefont {Iba\~{n}ez Azpiroz}, \citenamefont {Rousseau},
  \citenamefont {Eiguren},\ and\ \citenamefont {Bergara}}]{IbaNez-Azpiroz2014}%
  \BibitemOpen
  \bibfield  {author} {\bibinfo {author} {\bibfnamefont {J.}~\bibnamefont
  {Iba\~{n}ez Azpiroz}}, \bibinfo {author} {\bibfnamefont {B.}~\bibnamefont
  {Rousseau}}, \bibinfo {author} {\bibfnamefont {A.}~\bibnamefont {Eiguren}}, \
  and\ \bibinfo {author} {\bibfnamefont {A.}~\bibnamefont {Bergara}},\
  }\href@noop {} {\bibfield  {journal} {\bibinfo  {journal} {Phys. Rev. B}\
  }\textbf {\bibinfo {volume} {89}},\ \bibinfo {pages} {085102} (\bibinfo
  {year} {2014})}\BibitemShut {NoStop}%
\bibitem [{\citenamefont {Miao}\ and\ \citenamefont
  {Hoffman}(2015)}]{Miao2015}%
  \BibitemOpen
  \bibfield  {author} {\bibinfo {author} {\bibfnamefont {M.-s.}\ \bibnamefont
  {Miao}}\ and\ \bibinfo {author} {\bibfnamefont {R.}~\bibnamefont {Hoffman}},\
  }\href@noop {} {\bibfield  {journal} {\bibinfo  {journal} {J. Am. Chem.
  Soc.}\ }\textbf {\bibinfo {volume} {137}},\ \bibinfo {pages} {3631} (\bibinfo
  {year} {2015})}\BibitemShut {NoStop}%
\bibitem [{\citenamefont {Naumov}\ and\ \citenamefont
  {Hemley}(2015)}]{Naumov2015}%
  \BibitemOpen
  \bibfield  {author} {\bibinfo {author} {\bibfnamefont {I.~I.}\ \bibnamefont
  {Naumov}}\ and\ \bibinfo {author} {\bibfnamefont {R.~J.}\ \bibnamefont
  {Hemley}},\ }\href@noop {} {\bibfield  {journal} {\bibinfo  {journal} {Phys.
  Rev. Lett.}\ }\textbf {\bibinfo {volume} {114}},\ \bibinfo {pages} {156403}
  (\bibinfo {year} {2015})}\BibitemShut {NoStop}%
\bibitem [{\citenamefont {Gregoryanz}\ \emph {et~al.}(2005)\citenamefont
  {Gregoryanz}, \citenamefont {Degtyareva}, \citenamefont {Somayazulu},
  \citenamefont {Hemley},\ and\ \citenamefont {Mao}}]{Gregoryanz2005}%
  \BibitemOpen
  \bibfield  {author} {\bibinfo {author} {\bibfnamefont {E.}~\bibnamefont
  {Gregoryanz}}, \bibinfo {author} {\bibfnamefont {O.}~\bibnamefont
  {Degtyareva}}, \bibinfo {author} {\bibfnamefont {M.}~\bibnamefont
  {Somayazulu}}, \bibinfo {author} {\bibfnamefont {R.~J.}\ \bibnamefont
  {Hemley}}, \ and\ \bibinfo {author} {\bibfnamefont {H.-k.}\ \bibnamefont
  {Mao}},\ }\href {\doibase 10.1103/PhysRevLett.94.185502} {\bibfield
  {journal} {\bibinfo  {journal} {Phys. Rev. Lett.}\ }\textbf {\bibinfo
  {volume} {94}},\ \bibinfo {pages} {185502} (\bibinfo {year}
  {2005})}\BibitemShut {NoStop}%
\bibitem [{\citenamefont {Ko{\v{c}}i}\ \emph {et~al.}(2008)\citenamefont
  {Ko{\v{c}}i}, \citenamefont {Ahuja}, \citenamefont {Vitos},\ and\
  \citenamefont {Pinsook}}]{Koci2008}%
  \BibitemOpen
  \bibfield  {author} {\bibinfo {author} {\bibfnamefont {L.}~\bibnamefont
  {Ko{\v{c}}i}}, \bibinfo {author} {\bibfnamefont {R.}~\bibnamefont {Ahuja}},
  \bibinfo {author} {\bibfnamefont {L.}~\bibnamefont {Vitos}}, \ and\ \bibinfo
  {author} {\bibfnamefont {U.}~\bibnamefont {Pinsook}},\ }\href@noop {}
  {\bibfield  {journal} {\bibinfo  {journal} {Phys. Rev. B}\ }\textbf {\bibinfo
  {volume} {77}},\ \bibinfo {pages} {132101} (\bibinfo {year}
  {2008})}\BibitemShut {NoStop}%
\bibitem [{\citenamefont {Yamane}, \citenamefont {Shimojo},\ and\ \citenamefont
  {Hoshino}(2008)}]{Aki2008}%
  \BibitemOpen
  \bibfield  {author} {\bibinfo {author} {\bibfnamefont {A.}~\bibnamefont
  {Yamane}}, \bibinfo {author} {\bibfnamefont {F.}~\bibnamefont {Shimojo}}, \
  and\ \bibinfo {author} {\bibfnamefont {K.}~\bibnamefont {Hoshino}},\ }\href
  {\doibase 10.1143/JPSJ.77.064603} {\bibfield  {journal} {\bibinfo  {journal}
  {J. Phys. Soc. Japan}\ }\textbf {\bibinfo {volume} {77}},\ \bibinfo {pages}
  {064603} (\bibinfo {year} {2008})}\BibitemShut {NoStop}%
\bibitem [{\citenamefont {{Maksimov}}, \citenamefont {{Lepeshkin}},\ and\
  \citenamefont {{Magnitskaya}}(2011)}]{Maksimov2011}%
  \BibitemOpen
  \bibfield  {author} {\bibinfo {author} {\bibfnamefont {E.~G.}\ \bibnamefont
  {{Maksimov}}}, \bibinfo {author} {\bibfnamefont {S.~V.}\ \bibnamefont
  {{Lepeshkin}}}, \ and\ \bibinfo {author} {\bibfnamefont {M.~V.}\ \bibnamefont
  {{Magnitskaya}}},\ }\href {\doibase 10.1134/S1063774511040134} {\bibfield
  {journal} {\bibinfo  {journal} {Crystallogr. Rep.}\ }\textbf {\bibinfo
  {volume} {56}},\ \bibinfo {pages} {676} (\bibinfo {year} {2011})}\BibitemShut
  {NoStop}%
\bibitem [{\citenamefont {Eshet}\ \emph {et~al.}(2012)\citenamefont {Eshet},
  \citenamefont {Khaliullin}, \citenamefont {K{\"{u}}hne}, \citenamefont
  {Behler},\ and\ \citenamefont {Parrinello}}]{Eshet2012}%
  \BibitemOpen
  \bibfield  {author} {\bibinfo {author} {\bibfnamefont {H.}~\bibnamefont
  {Eshet}}, \bibinfo {author} {\bibfnamefont {R.~Z.}\ \bibnamefont
  {Khaliullin}}, \bibinfo {author} {\bibfnamefont {T.~D.}\ \bibnamefont
  {K{\"{u}}hne}}, \bibinfo {author} {\bibfnamefont {J.}~\bibnamefont {Behler}},
  \ and\ \bibinfo {author} {\bibfnamefont {M.}~\bibnamefont {Parrinello}},\
  }\href@noop {} {\bibfield  {journal} {\bibinfo  {journal} {Phys. Rev. Lett.}\
  }\textbf {\bibinfo {volume} {108}},\ \bibinfo {pages} {115701} (\bibinfo
  {year} {2012})}\BibitemShut {NoStop}%
\bibitem [{\citenamefont {Gonz\'{a}lez}\ and\ \citenamefont
  {Gonz\'{a}lez}(2015)}]{Gonzalez2015}%
  \BibitemOpen
  \bibfield  {author} {\bibinfo {author} {\bibfnamefont {D.~J.}\ \bibnamefont
  {Gonz\'{a}lez}}\ and\ \bibinfo {author} {\bibfnamefont {L.~E.}\ \bibnamefont
  {Gonz\'{a}lez}},\ }\href
  {http://scitation.aip.org/content/aip/proceeding/aipcp/10.1063/1.4928259}
  {\bibfield  {journal} {\bibinfo  {journal} {AIP Conf. Proc.}\ }\textbf
  {\bibinfo {volume} {1673}},\ \bibinfo {eid} {020005} (\bibinfo {year}
  {2015})}\BibitemShut {NoStop}%
\bibitem [{\citenamefont {Li}\ \emph {et~al.}(2015)\citenamefont {Li},
  \citenamefont {Wang}, \citenamefont {Pickard}, \citenamefont {Needs},
  \citenamefont {Wang},\ and\ \citenamefont {Ma}}]{Li2015}%
  \BibitemOpen
  \bibfield  {author} {\bibinfo {author} {\bibfnamefont {Y.}~\bibnamefont
  {Li}}, \bibinfo {author} {\bibfnamefont {Y.}~\bibnamefont {Wang}}, \bibinfo
  {author} {\bibfnamefont {C.~J.}\ \bibnamefont {Pickard}}, \bibinfo {author}
  {\bibfnamefont {R.~J.}\ \bibnamefont {Needs}}, \bibinfo {author}
  {\bibfnamefont {Y.}~\bibnamefont {Wang}}, \ and\ \bibinfo {author}
  {\bibfnamefont {Y.}~\bibnamefont {Ma}},\ }\href@noop {} {\bibfield  {journal}
  {\bibinfo  {journal} {Phys. Rev. Lett.}\ }\textbf {\bibinfo {volume} {114}},\
  \bibinfo {pages} {125501} (\bibinfo {year} {2015})}\BibitemShut {NoStop}%
\bibitem [{\citenamefont {Swift}\ \emph {et~al.}(2012)\citenamefont {Swift},
  \citenamefont {Hawreliak}, \citenamefont {Braun}, \citenamefont {Kritcher},
  \citenamefont {Glenzer}, \citenamefont {Collins}, \citenamefont {Rothman},
  \citenamefont {Chapman},\ and\ \citenamefont {Rose}}]{Swift2012}%
  \BibitemOpen
  \bibfield  {author} {\bibinfo {author} {\bibfnamefont {D.}~\bibnamefont
  {Swift}}, \bibinfo {author} {\bibfnamefont {J.}~\bibnamefont {Hawreliak}},
  \bibinfo {author} {\bibfnamefont {D.}~\bibnamefont {Braun}}, \bibinfo
  {author} {\bibfnamefont {A.}~\bibnamefont {Kritcher}}, \bibinfo {author}
  {\bibfnamefont {S.}~\bibnamefont {Glenzer}}, \bibinfo {author} {\bibfnamefont
  {G.~W.}\ \bibnamefont {Collins}}, \bibinfo {author} {\bibfnamefont
  {S.}~\bibnamefont {Rothman}}, \bibinfo {author} {\bibfnamefont
  {D.}~\bibnamefont {Chapman}}, \ and\ \bibinfo {author} {\bibfnamefont
  {S.}~\bibnamefont {Rose}},\ }\href {\doibase
  http://dx.doi.org/10.1063/1.3686321} {\bibfield  {journal} {\bibinfo
  {journal} {AIP Conf. Proc.}\ }\textbf {\bibinfo {volume} {1426}},\ \bibinfo
  {pages} {477} (\bibinfo {year} {2012})}\BibitemShut {NoStop}%
\bibitem [{\citenamefont {Kritcher}\ \emph {et~al.}(2016)\citenamefont
  {Kritcher}, \citenamefont {Doeppner}, \citenamefont {Swift}, \citenamefont
  {Hawreliak}, \citenamefont {Nilsen}, \citenamefont {Hammer}, \citenamefont
  {Bachmann}, \citenamefont {Collins}, \citenamefont {Landen}, \citenamefont
  {Keane}, \citenamefont {Glenzer}, \citenamefont {Rothman}, \citenamefont
  {Chapman}, \citenamefont {Kraus},\ and\ \citenamefont
  {Falcone}}]{Kritcher2016}%
  \BibitemOpen
  \bibfield  {author} {\bibinfo {author} {\bibfnamefont {A.~L.}\ \bibnamefont
  {Kritcher}}, \bibinfo {author} {\bibfnamefont {T.}~\bibnamefont {Doeppner}},
  \bibinfo {author} {\bibfnamefont {D.}~\bibnamefont {Swift}}, \bibinfo
  {author} {\bibfnamefont {J.}~\bibnamefont {Hawreliak}}, \bibinfo {author}
  {\bibfnamefont {J.}~\bibnamefont {Nilsen}}, \bibinfo {author} {\bibfnamefont
  {J.}~\bibnamefont {Hammer}}, \bibinfo {author} {\bibfnamefont
  {B.}~\bibnamefont {Bachmann}}, \bibinfo {author} {\bibfnamefont
  {G.}~\bibnamefont {Collins}}, \bibinfo {author} {\bibfnamefont
  {O.}~\bibnamefont {Landen}}, \bibinfo {author} {\bibfnamefont
  {C.}~\bibnamefont {Keane}}, \bibinfo {author} {\bibfnamefont
  {S.}~\bibnamefont {Glenzer}}, \bibinfo {author} {\bibfnamefont
  {S.}~\bibnamefont {Rothman}}, \bibinfo {author} {\bibfnamefont
  {D.}~\bibnamefont {Chapman}}, \bibinfo {author} {\bibfnamefont
  {D.}~\bibnamefont {Kraus}}, \ and\ \bibinfo {author} {\bibfnamefont
  {R.}~\bibnamefont {Falcone}},\ }\href@noop {} {\bibfield  {journal} {\bibinfo
   {journal} {J. Phys. Conf. Ser.}\ }\textbf {\bibinfo {volume} {688}},\
  \bibinfo {pages} {012055} (\bibinfo {year} {2016})}\BibitemShut {NoStop}%
\bibitem [{\citenamefont {Golyshev}\ \emph {et~al.}(2011)\citenamefont
  {Golyshev}, \citenamefont {Shakhray}, \citenamefont {Kim}, \citenamefont
  {Molodets},\ and\ \citenamefont {Fortov}}]{Golyshev2011}%
  \BibitemOpen
  \bibfield  {author} {\bibinfo {author} {\bibfnamefont {A.~A.}\ \bibnamefont
  {Golyshev}}, \bibinfo {author} {\bibfnamefont {D.~V.}\ \bibnamefont
  {Shakhray}}, \bibinfo {author} {\bibfnamefont {V.~V.}\ \bibnamefont {Kim}},
  \bibinfo {author} {\bibfnamefont {A.~M.}\ \bibnamefont {Molodets}}, \ and\
  \bibinfo {author} {\bibfnamefont {V.~E.}\ \bibnamefont {Fortov}},\ }\href
  {\doibase 10.1103/PhysRevB.83.094114} {\bibfield  {journal} {\bibinfo
  {journal} {Phys. Rev. B}\ }\textbf {\bibinfo {volume} {83}},\ \bibinfo
  {pages} {094114} (\bibinfo {year} {2011})}\BibitemShut {NoStop}%
\bibitem [{\citenamefont {Rice}(1965)}]{Rice1965}%
  \BibitemOpen
  \bibfield  {author} {\bibinfo {author} {\bibfnamefont {M.~H.}\ \bibnamefont
  {Rice}},\ }\href@noop {} {\bibfield  {journal} {\bibinfo  {journal} {J. Phys.
  Chem. Solids}\ }\textbf {\bibinfo {volume} {26}},\ \bibinfo {pages} {483}
  (\bibinfo {year} {1965})}\BibitemShut {NoStop}%
\bibitem [{\citenamefont {Bakanova}, \citenamefont {Dudoladov},\ and\
  \citenamefont {Trunin}(1965)}]{Bakanova1965}%
  \BibitemOpen
  \bibfield  {author} {\bibinfo {author} {\bibfnamefont {A.}~\bibnamefont
  {Bakanova}}, \bibinfo {author} {\bibfnamefont {I.~P.}\ \bibnamefont
  {Dudoladov}}, \ and\ \bibinfo {author} {\bibfnamefont {R.~F.}\ \bibnamefont
  {Trunin}},\ }\href@noop {} {\bibfield  {journal} {\bibinfo  {journal} {Sov.
  Phys. Solid State}\ }\textbf {\bibinfo {volume} {7}},\ \bibinfo {pages}
  {1307} (\bibinfo {year} {1965})}\BibitemShut {NoStop}%
\bibitem [{Mar()}]{MarshLASL1980}%
  \BibitemOpen
  \href@noop {} {}\bibinfo {note} {LASL Shock Hugoniot Data, edited by S. P.
  Marsh (University of California Press, Berkeley, 1980).}\BibitemShut {Stop}%
\bibitem [{\citenamefont {Ehrenfeld}, \citenamefont {Krimsky},\ and\
  \citenamefont {Selvitella}(1966)}]{Ehrenfeld1966analyticHug}%
  \BibitemOpen
  \bibfield  {author} {\bibinfo {author} {\bibfnamefont {J.}~\bibnamefont
  {Ehrenfeld}}, \bibinfo {author} {\bibfnamefont {S.}~\bibnamefont {Krimsky}},
  \ and\ \bibinfo {author} {\bibfnamefont {J.}~\bibnamefont {Selvitella}},\
  }\href@noop {} {\bibfield  {journal} {\bibinfo  {journal} {J. Appl. Phys.}\
  }\textbf {\bibinfo {volume} {37}},\ \bibinfo {pages} {4737} (\bibinfo {year}
  {1966})}\BibitemShut {NoStop}%
\bibitem [{\citenamefont {Hickey}(1967)}]{Hickey1967comment}%
  \BibitemOpen
  \bibfield  {author} {\bibinfo {author} {\bibfnamefont {D.~P.}\ \bibnamefont
  {Hickey}},\ }\href@noop {} {\bibfield  {journal} {\bibinfo  {journal} {J.
  Appl. Phys.}\ }\textbf {\bibinfo {volume} {38}},\ \bibinfo {pages} {4080}
  (\bibinfo {year} {1967})}\BibitemShut {NoStop}%
\bibitem [{\citenamefont {Pastine}(1967)}]{Pastine1967}%
  \BibitemOpen
  \bibfield  {author} {\bibinfo {author} {\bibfnamefont {D.~J.}\ \bibnamefont
  {Pastine}},\ }\href {\doibase 10.1103/PhysRevLett.18.1187} {\bibfield
  {journal} {\bibinfo  {journal} {Phys. Rev. Lett.}\ }\textbf {\bibinfo
  {volume} {18}},\ \bibinfo {pages} {1187} (\bibinfo {year}
  {1967})}\BibitemShut {NoStop}%
\bibitem [{\citenamefont {Pastine}(1968)}]{Pastine1968}%
  \BibitemOpen
  \bibfield  {author} {\bibinfo {author} {\bibfnamefont {D.~J.}\ \bibnamefont
  {Pastine}},\ }\href {\doibase 10.1103/PhysRev.166.703} {\bibfield  {journal}
  {\bibinfo  {journal} {Phys. Rev.}\ }\textbf {\bibinfo {volume} {166}},\
  \bibinfo {pages} {703} (\bibinfo {year} {1968})}\BibitemShut {NoStop}%
\bibitem [{\citenamefont {Khishchenko}(2008)}]{Khishchenko2008}%
  \BibitemOpen
  \bibfield  {author} {\bibinfo {author} {\bibfnamefont {K.~V.}\ \bibnamefont
  {Khishchenko}},\ }\href {http://stacks.iop.org/1742-6596/98/i=3/a=032023}
  {\bibfield  {journal} {\bibinfo  {journal} {J. Phys. Conf. Ser.}\ }\textbf
  {\bibinfo {volume} {98}},\ \bibinfo {pages} {032023} (\bibinfo {year}
  {2008})}\BibitemShut {NoStop}%
\bibitem [{\citenamefont {Ross}(1980)}]{Ross1980}%
  \BibitemOpen
  \bibfield  {author} {\bibinfo {author} {\bibfnamefont {M.}~\bibnamefont
  {Ross}},\ }\href {\doibase 10.1103/PhysRevB.21.3140} {\bibfield  {journal}
  {\bibinfo  {journal} {Phys. Rev. B}\ }\textbf {\bibinfo {volume} {21}},\
  \bibinfo {pages} {3140} (\bibinfo {year} {1980})}\BibitemShut {NoStop}%
\bibitem [{\citenamefont {Young}\ and\ \citenamefont {Ross}(1984)}]{Young1984}%
  \BibitemOpen
  \bibfield  {author} {\bibinfo {author} {\bibfnamefont {D.~A.}\ \bibnamefont
  {Young}}\ and\ \bibinfo {author} {\bibfnamefont {M.}~\bibnamefont {Ross}},\
  }\href@noop {} {\bibfield  {journal} {\bibinfo  {journal} {Phys. Rev. B}\
  }\textbf {\bibinfo {volume} {29}},\ \bibinfo {pages} {682} (\bibinfo {year}
  {1984})}\BibitemShut {NoStop}%
\bibitem [{\citenamefont {Belashchenko}(2009)}]{Belashchenko2009}%
  \BibitemOpen
  \bibfield  {author} {\bibinfo {author} {\bibfnamefont {D.~K.}\ \bibnamefont
  {Belashchenko}},\ }\href@noop {} {\bibfield  {journal} {\bibinfo  {journal}
  {High Temp.}\ }\textbf {\bibinfo {volume} {47}},\ \bibinfo {pages} {494}
  (\bibinfo {year} {2009})}\BibitemShut {NoStop}%
\bibitem [{\citenamefont
  {Belashchenko}(2013{\natexlab{a}})}]{Belashchenko2013a}%
  \BibitemOpen
  \bibfield  {author} {\bibinfo {author} {\bibfnamefont {D.~K.}\ \bibnamefont
  {Belashchenko}},\ }\href@noop {} {\bibfield  {journal} {\bibinfo  {journal}
  {High Temp.}\ }\textbf {\bibinfo {volume} {51}},\ \bibinfo {pages} {626}
  (\bibinfo {year} {2013}{\natexlab{a}})}\BibitemShut {NoStop}%
\bibitem [{\citenamefont
  {Belashchenko}(2013{\natexlab{b}})}]{Belashchenko2013b}%
  \BibitemOpen
  \bibfield  {author} {\bibinfo {author} {\bibfnamefont {D.~K.}\ \bibnamefont
  {Belashchenko}},\ }\href {http://stacks.iop.org/1063-7869/56/i=12/a=1176}
  {\bibfield  {journal} {\bibinfo  {journal} {Phys. Usp.}\ }\textbf {\bibinfo
  {volume} {56}},\ \bibinfo {pages} {1176} (\bibinfo {year}
  {2013}{\natexlab{b}})}\BibitemShut {NoStop}%
\bibitem [{\citenamefont {Pierleoni}\ \emph {et~al.}(1994)\citenamefont
  {Pierleoni}, \citenamefont {Ceperley}, \citenamefont {Bernu},\ and\
  \citenamefont {Magro}}]{PhysRevLett.73.2145}%
  \BibitemOpen
  \bibfield  {author} {\bibinfo {author} {\bibfnamefont {C.}~\bibnamefont
  {Pierleoni}}, \bibinfo {author} {\bibfnamefont {D.~M.}\ \bibnamefont
  {Ceperley}}, \bibinfo {author} {\bibfnamefont {B.}~\bibnamefont {Bernu}}, \
  and\ \bibinfo {author} {\bibfnamefont {W.~R.}\ \bibnamefont {Magro}},\ }\href
  {\doibase 10.1103/PhysRevLett.73.2145} {\bibfield  {journal} {\bibinfo
  {journal} {Phys. Rev. Lett.}\ }\textbf {\bibinfo {volume} {73}},\ \bibinfo
  {pages} {2145} (\bibinfo {year} {1994})}\BibitemShut {NoStop}%
\bibitem [{\citenamefont {Driver}\ and\ \citenamefont
  {Militzer}(2012)}]{Driver2012}%
  \BibitemOpen
  \bibfield  {author} {\bibinfo {author} {\bibfnamefont {K.~P.}\ \bibnamefont
  {Driver}}\ and\ \bibinfo {author} {\bibfnamefont {B.}~\bibnamefont
  {Militzer}},\ }\href@noop {} {\bibfield  {journal} {\bibinfo  {journal}
  {Phys. Rev. Lett.}\ }\textbf {\bibinfo {volume} {108}},\ \bibinfo {pages}
  {115502} (\bibinfo {year} {2012})}\BibitemShut {NoStop}%
\bibitem [{\citenamefont {Militzer}\ and\ \citenamefont
  {Driver}(2015)}]{Militzer2015}%
  \BibitemOpen
  \bibfield  {author} {\bibinfo {author} {\bibfnamefont {B.}~\bibnamefont
  {Militzer}}\ and\ \bibinfo {author} {\bibfnamefont {K.~P.}\ \bibnamefont
  {Driver}},\ }\href@noop {} {\bibfield  {journal} {\bibinfo  {journal} {Phys.
  Rev. Lett.}\ }\textbf {\bibinfo {volume} {115}},\ \bibinfo {pages} {176403}
  (\bibinfo {year} {2015})}\BibitemShut {NoStop}%
\bibitem [{\citenamefont {Hu}\ \emph {et~al.}(2016)\citenamefont {Hu},
  \citenamefont {Militzer}, \citenamefont {Collins}, \citenamefont {Driver},\
  and\ \citenamefont {Kress}}]{PhysRevB.94.094109}%
  \BibitemOpen
  \bibfield  {author} {\bibinfo {author} {\bibfnamefont {S.~X.}\ \bibnamefont
  {Hu}}, \bibinfo {author} {\bibfnamefont {B.}~\bibnamefont {Militzer}},
  \bibinfo {author} {\bibfnamefont {L.~A.}\ \bibnamefont {Collins}}, \bibinfo
  {author} {\bibfnamefont {K.~P.}\ \bibnamefont {Driver}}, \ and\ \bibinfo
  {author} {\bibfnamefont {J.~D.}\ \bibnamefont {Kress}},\ }\href {\doibase
  10.1103/PhysRevB.94.094109} {\bibfield  {journal} {\bibinfo  {journal} {Phys.
  Rev. B}\ }\textbf {\bibinfo {volume} {94}},\ \bibinfo {pages} {094109}
  (\bibinfo {year} {2016})}\BibitemShut {NoStop}%
\bibitem [{\citenamefont {Lambert}\ and\ \citenamefont
  {Recoules}(2012)}]{PhysRevE.86.026405}%
  \BibitemOpen
  \bibfield  {author} {\bibinfo {author} {\bibfnamefont {F.}~\bibnamefont
  {Lambert}}\ and\ \bibinfo {author} {\bibfnamefont {V.}~\bibnamefont
  {Recoules}},\ }\href {\doibase 10.1103/PhysRevE.86.026405} {\bibfield
  {journal} {\bibinfo  {journal} {Phys. Rev. E}\ }\textbf {\bibinfo {volume}
  {86}},\ \bibinfo {pages} {026405} (\bibinfo {year} {2012})}\BibitemShut
  {NoStop}%
\bibitem [{\citenamefont {Gao}\ \emph {et~al.}(2016)\citenamefont {Gao},
  \citenamefont {Zhang}, \citenamefont {Kang}, \citenamefont {Wang},
  \citenamefont {Zhang},\ and\ \citenamefont {He}}]{PhysRevB.94.205115}%
  \BibitemOpen
  \bibfield  {author} {\bibinfo {author} {\bibfnamefont {C.}~\bibnamefont
  {Gao}}, \bibinfo {author} {\bibfnamefont {S.}~\bibnamefont {Zhang}}, \bibinfo
  {author} {\bibfnamefont {W.}~\bibnamefont {Kang}}, \bibinfo {author}
  {\bibfnamefont {C.}~\bibnamefont {Wang}}, \bibinfo {author} {\bibfnamefont
  {P.}~\bibnamefont {Zhang}}, \ and\ \bibinfo {author} {\bibfnamefont {X.~T.}\
  \bibnamefont {He}},\ }\href {\doibase 10.1103/PhysRevB.94.205115} {\bibfield
  {journal} {\bibinfo  {journal} {Phys. Rev. B}\ }\textbf {\bibinfo {volume}
  {94}},\ \bibinfo {pages} {205115} (\bibinfo {year} {2016})}\BibitemShut
  {NoStop}%
\bibitem [{\citenamefont {Zhang}\ \emph
  {et~al.}(2016{\natexlab{a}})\citenamefont {Zhang}, \citenamefont {Wang},
  \citenamefont {Kang}, \citenamefont {Zhang},\ and\ \citenamefont
  {He}}]{ZhangExtendedDFT2016}%
  \BibitemOpen
  \bibfield  {author} {\bibinfo {author} {\bibfnamefont {S.}~\bibnamefont
  {Zhang}}, \bibinfo {author} {\bibfnamefont {H.}~\bibnamefont {Wang}},
  \bibinfo {author} {\bibfnamefont {W.}~\bibnamefont {Kang}}, \bibinfo {author}
  {\bibfnamefont {P.}~\bibnamefont {Zhang}}, \ and\ \bibinfo {author}
  {\bibfnamefont {X.~T.}\ \bibnamefont {He}},\ }\href
  {http://scitation.aip.org/content/aip/journal/pop/23/4/10.1063/1.4947212}
  {\bibfield  {journal} {\bibinfo  {journal} {Phys. Plasmas}\ }\textbf
  {\bibinfo {volume} {23}},\ \bibinfo {eid} {042707} (\bibinfo {year}
  {2016}{\natexlab{a}})}\BibitemShut {NoStop}%
\bibitem [{\citenamefont {Militzer}\ and\ \citenamefont
  {Ceperley}(2001)}]{PhysRevE.63.066404}%
  \BibitemOpen
  \bibfield  {author} {\bibinfo {author} {\bibfnamefont {B.}~\bibnamefont
  {Militzer}}\ and\ \bibinfo {author} {\bibfnamefont {D.~M.}\ \bibnamefont
  {Ceperley}},\ }\href {\doibase 10.1103/PhysRevE.63.066404} {\bibfield
  {journal} {\bibinfo  {journal} {Phys. Rev. E}\ }\textbf {\bibinfo {volume}
  {63}},\ \bibinfo {pages} {066404} (\bibinfo {year} {2001})}\BibitemShut
  {NoStop}%
\bibitem [{\citenamefont {Militzer}(2009)}]{Militzer2009}%
  \BibitemOpen
  \bibfield  {author} {\bibinfo {author} {\bibfnamefont {B.}~\bibnamefont
  {Militzer}},\ }\href@noop {} {\bibfield  {journal} {\bibinfo  {journal}
  {Phys. Rev. B}\ }\textbf {\bibinfo {volume} {79}},\ \bibinfo {pages} {155105}
  (\bibinfo {year} {2009})}\BibitemShut {NoStop}%
\bibitem [{\citenamefont {Benedict}\ \emph {et~al.}(2014)\citenamefont
  {Benedict}, \citenamefont {Driver}, \citenamefont {Hamel}, \citenamefont
  {Militzer}, \citenamefont {Qi}, \citenamefont {Correa}, \citenamefont
  {Saul},\ and\ \citenamefont {Schwegler}}]{Benedict2014C}%
  \BibitemOpen
  \bibfield  {author} {\bibinfo {author} {\bibfnamefont {L.~X.}\ \bibnamefont
  {Benedict}}, \bibinfo {author} {\bibfnamefont {K.~P.}\ \bibnamefont
  {Driver}}, \bibinfo {author} {\bibfnamefont {S.}~\bibnamefont {Hamel}},
  \bibinfo {author} {\bibfnamefont {B.}~\bibnamefont {Militzer}}, \bibinfo
  {author} {\bibfnamefont {T.}~\bibnamefont {Qi}}, \bibinfo {author}
  {\bibfnamefont {A.~A.}\ \bibnamefont {Correa}}, \bibinfo {author}
  {\bibfnamefont {A.}~\bibnamefont {Saul}}, \ and\ \bibinfo {author}
  {\bibfnamefont {E.}~\bibnamefont {Schwegler}},\ }\href@noop {} {\bibfield
  {journal} {\bibinfo  {journal} {Phys. Rev. B}\ }\textbf {\bibinfo {volume}
  {89}},\ \bibinfo {pages} {224109} (\bibinfo {year} {2014})}\BibitemShut
  {NoStop}%
\bibitem [{\citenamefont {Driver}\ and\ \citenamefont
  {Militzer}(2015)}]{Driver2015Neon}%
  \BibitemOpen
  \bibfield  {author} {\bibinfo {author} {\bibfnamefont {K.~P.}\ \bibnamefont
  {Driver}}\ and\ \bibinfo {author} {\bibfnamefont {B.}~\bibnamefont
  {Militzer}},\ }\href@noop {} {\bibfield  {journal} {\bibinfo  {journal}
  {Phys. Rev. B}\ }\textbf {\bibinfo {volume} {91}},\ \bibinfo {pages} {045103}
  (\bibinfo {year} {2015})}\BibitemShut {NoStop}%
\bibitem [{\citenamefont {Driver}\ \emph {et~al.}(2015)\citenamefont {Driver},
  \citenamefont {Soubiran}, \citenamefont {Zhang},\ and\ \citenamefont
  {Militzer}}]{Driver2015Oxygen}%
  \BibitemOpen
  \bibfield  {author} {\bibinfo {author} {\bibfnamefont {K.~P.}\ \bibnamefont
  {Driver}}, \bibinfo {author} {\bibfnamefont {F.}~\bibnamefont {Soubiran}},
  \bibinfo {author} {\bibfnamefont {S.}~\bibnamefont {Zhang}}, \ and\ \bibinfo
  {author} {\bibfnamefont {B.}~\bibnamefont {Militzer}},\ }\href@noop {}
  {\bibfield  {journal} {\bibinfo  {journal} {J. Chem. Phys.}\ }\textbf
  {\bibinfo {volume} {143}},\ \bibinfo {pages} {164507} (\bibinfo {year}
  {2015})}\BibitemShut {NoStop}%
\bibitem [{\citenamefont {Driver}\ and\ \citenamefont
  {Militzer}(2016)}]{Driver2016Nitrogen}%
  \BibitemOpen
  \bibfield  {author} {\bibinfo {author} {\bibfnamefont {K.~P.}\ \bibnamefont
  {Driver}}\ and\ \bibinfo {author} {\bibfnamefont {B.}~\bibnamefont
  {Militzer}},\ }\href@noop {} {\bibfield  {journal} {\bibinfo  {journal}
  {Phys. Rev. B}\ }\textbf {\bibinfo {volume} {93}},\ \bibinfo {pages} {064101}
  (\bibinfo {year} {2016})}\BibitemShut {NoStop}%
\bibitem [{\citenamefont {Zhang}\ \emph
  {et~al.}(2016{\natexlab{b}})\citenamefont {Zhang}, \citenamefont {Driver},
  \citenamefont {Soubiran},\ and\ \citenamefont {Militzer}}]{Zhang2016b}%
  \BibitemOpen
  \bibfield  {author} {\bibinfo {author} {\bibfnamefont {S.}~\bibnamefont
  {Zhang}}, \bibinfo {author} {\bibfnamefont {K.~P.}\ \bibnamefont {Driver}},
  \bibinfo {author} {\bibfnamefont {F.}~\bibnamefont {Soubiran}}, \ and\
  \bibinfo {author} {\bibfnamefont {B.}~\bibnamefont {Militzer}},\ }\href
  {\doibase 10.1016/j.hedp.2016.09.004} {\bibfield  {journal} {\bibinfo
  {journal} {High Energ. Dens. Phys.}\ }\textbf {\bibinfo {volume} {21}},\
  \bibinfo {pages} {16} (\bibinfo {year} {2016}{\natexlab{b}})}\BibitemShut
  {NoStop}%
\bibitem [{\citenamefont {Debye}\ and\ \citenamefont
  {Huckel}(1923)}]{DebyeHuckel}%
  \BibitemOpen
  \bibfield  {author} {\bibinfo {author} {\bibfnamefont {P.}~\bibnamefont
  {Debye}}\ and\ \bibinfo {author} {\bibfnamefont {E.}~\bibnamefont {Huckel}},\
  }\href@noop {} {\bibfield  {journal} {\bibinfo  {journal} {Phys. Z.}\
  }\textbf {\bibinfo {volume} {24}},\ \bibinfo {pages} {185} (\bibinfo {year}
  {1923})}\BibitemShut {NoStop}%
\bibitem [{\citenamefont {Sjostrom}\ and\ \citenamefont
  {Daligault}(2014)}]{Sjostrom2014}%
  \BibitemOpen
  \bibfield  {author} {\bibinfo {author} {\bibfnamefont {T.}~\bibnamefont
  {Sjostrom}}\ and\ \bibinfo {author} {\bibfnamefont {J.}~\bibnamefont
  {Daligault}},\ }\href@noop {} {\bibfield  {journal} {\bibinfo  {journal}
  {Phys. Rev. Lett.}\ }\textbf {\bibinfo {volume} {113}},\ \bibinfo {pages}
  {155006} (\bibinfo {year} {2014})}\BibitemShut {NoStop}%
\bibitem [{\citenamefont {Karasiev}, \citenamefont {Calder\'{\i}n},\ and\
  \citenamefont {Trickey}(2016)}]{Karasiev2016}%
  \BibitemOpen
  \bibfield  {author} {\bibinfo {author} {\bibfnamefont {V.~V.}\ \bibnamefont
  {Karasiev}}, \bibinfo {author} {\bibfnamefont {L.}~\bibnamefont
  {Calder\'{\i}n}}, \ and\ \bibinfo {author} {\bibfnamefont {S.~B.}\
  \bibnamefont {Trickey}},\ }\href@noop {} {\bibfield  {journal} {\bibinfo
  {journal} {Phys. Rev. E}\ }\textbf {\bibinfo {volume} {93}},\ \bibinfo
  {pages} {063207} (\bibinfo {year} {2016})}\BibitemShut {NoStop}%
\bibitem [{\citenamefont {Feynman}(1953)}]{Feynman1953}%
  \BibitemOpen
  \bibfield  {author} {\bibinfo {author} {\bibfnamefont {R.~P.}\ \bibnamefont
  {Feynman}},\ }\href {\doibase 10.1103/PhysRev.91.1291} {\bibfield  {journal}
  {\bibinfo  {journal} {Phys. Rev.}\ }\textbf {\bibinfo {volume} {91}},\
  \bibinfo {pages} {1291} (\bibinfo {year} {1953})}\BibitemShut {NoStop}%
\bibitem [{\citenamefont {Magro}\ \emph {et~al.}(1996)\citenamefont {Magro},
  \citenamefont {Ceperley}, \citenamefont {Pierleoni},\ and\ \citenamefont
  {Bernu}}]{PhysRevLett.76.1240}%
  \BibitemOpen
  \bibfield  {author} {\bibinfo {author} {\bibfnamefont {W.~R.}\ \bibnamefont
  {Magro}}, \bibinfo {author} {\bibfnamefont {D.~M.}\ \bibnamefont {Ceperley}},
  \bibinfo {author} {\bibfnamefont {C.}~\bibnamefont {Pierleoni}}, \ and\
  \bibinfo {author} {\bibfnamefont {B.}~\bibnamefont {Bernu}},\ }\href
  {\doibase 10.1103/PhysRevLett.76.1240} {\bibfield  {journal} {\bibinfo
  {journal} {Phys. Rev. Lett.}\ }\textbf {\bibinfo {volume} {76}},\ \bibinfo
  {pages} {1240} (\bibinfo {year} {1996})}\BibitemShut {NoStop}%
\bibitem [{\citenamefont {Militzer}\ \emph {et~al.}(2001)\citenamefont
  {Militzer}, \citenamefont {Ceperley}, \citenamefont {Kress}, \citenamefont
  {Johnson}, \citenamefont {Collins},\ and\ \citenamefont
  {Mazevet}}]{PhysRevLett.87.275502}%
  \BibitemOpen
  \bibfield  {author} {\bibinfo {author} {\bibfnamefont {B.}~\bibnamefont
  {Militzer}}, \bibinfo {author} {\bibfnamefont {D.~M.}\ \bibnamefont
  {Ceperley}}, \bibinfo {author} {\bibfnamefont {J.~D.}\ \bibnamefont {Kress}},
  \bibinfo {author} {\bibfnamefont {J.~D.}\ \bibnamefont {Johnson}}, \bibinfo
  {author} {\bibfnamefont {L.~A.}\ \bibnamefont {Collins}}, \ and\ \bibinfo
  {author} {\bibfnamefont {S.}~\bibnamefont {Mazevet}},\ }\href {\doibase
  10.1103/PhysRevLett.87.275502} {\bibfield  {journal} {\bibinfo  {journal}
  {Phys. Rev. Lett.}\ }\textbf {\bibinfo {volume} {87}},\ \bibinfo {pages}
  {275502} (\bibinfo {year} {2001})}\BibitemShut {NoStop}%
\bibitem [{\citenamefont {Hu}\ \emph {et~al.}(2010)\citenamefont {Hu},
  \citenamefont {Militzer}, \citenamefont {Goncharov},\ and\ \citenamefont
  {Skupsky}}]{PhysRevLett.104.235003}%
  \BibitemOpen
  \bibfield  {author} {\bibinfo {author} {\bibfnamefont {S.~X.}\ \bibnamefont
  {Hu}}, \bibinfo {author} {\bibfnamefont {B.}~\bibnamefont {Militzer}},
  \bibinfo {author} {\bibfnamefont {V.~N.}\ \bibnamefont {Goncharov}}, \ and\
  \bibinfo {author} {\bibfnamefont {S.}~\bibnamefont {Skupsky}},\ }\href
  {\doibase 10.1103/PhysRevLett.104.235003} {\bibfield  {journal} {\bibinfo
  {journal} {Phys. Rev. Lett.}\ }\textbf {\bibinfo {volume} {104}},\ \bibinfo
  {pages} {235003} (\bibinfo {year} {2010})}\BibitemShut {NoStop}%
\bibitem [{\citenamefont {Militzer}\ and\ \citenamefont
  {Ceperley}(2000)}]{PhysRevLett.85.1890}%
  \BibitemOpen
  \bibfield  {author} {\bibinfo {author} {\bibfnamefont {B.}~\bibnamefont
  {Militzer}}\ and\ \bibinfo {author} {\bibfnamefont {D.~M.}\ \bibnamefont
  {Ceperley}},\ }\href@noop {} {\bibfield  {journal} {\bibinfo  {journal}
  {Phys. Rev. Lett.}\ }\textbf {\bibinfo {volume} {85}},\ \bibinfo {pages}
  {1890} (\bibinfo {year} {2000})}\BibitemShut {NoStop}%
\bibitem [{\citenamefont {Hu}\ \emph {et~al.}(2011)\citenamefont {Hu},
  \citenamefont {Militzer}, \citenamefont {Goncharov},\ and\ \citenamefont
  {Skupsky}}]{PhysRevB.84.224109}%
  \BibitemOpen
  \bibfield  {author} {\bibinfo {author} {\bibfnamefont {S.~X.}\ \bibnamefont
  {Hu}}, \bibinfo {author} {\bibfnamefont {B.}~\bibnamefont {Militzer}},
  \bibinfo {author} {\bibfnamefont {V.~N.}\ \bibnamefont {Goncharov}}, \ and\
  \bibinfo {author} {\bibfnamefont {S.}~\bibnamefont {Skupsky}},\ }\href
  {\doibase 10.1103/PhysRevB.84.224109} {\bibfield  {journal} {\bibinfo
  {journal} {Phys. Rev. B}\ }\textbf {\bibinfo {volume} {84}},\ \bibinfo
  {pages} {224109} (\bibinfo {year} {2011})}\BibitemShut {NoStop}%
\bibitem [{\citenamefont {Militzer}, \citenamefont {Magro},\ and\ \citenamefont
  {Ceperley}(1999)}]{CTPP:CTPP2150390137}%
  \BibitemOpen
  \bibfield  {author} {\bibinfo {author} {\bibfnamefont {B.}~\bibnamefont
  {Militzer}}, \bibinfo {author} {\bibfnamefont {W.}~\bibnamefont {Magro}}, \
  and\ \bibinfo {author} {\bibfnamefont {D.}~\bibnamefont {Ceperley}},\ }\href
  {\doibase 10.1002/ctpp.2150390137} {\bibfield  {journal} {\bibinfo  {journal}
  {Contrib. Plasm. Phys.}\ }\textbf {\bibinfo {volume} {39}},\ \bibinfo {pages}
  {151} (\bibinfo {year} {1999})}\BibitemShut {NoStop}%
\bibitem [{\citenamefont {Militzer}\ and\ \citenamefont
  {Graham}(2006)}]{Militzer20062136}%
  \BibitemOpen
  \bibfield  {author} {\bibinfo {author} {\bibfnamefont {B.}~\bibnamefont
  {Militzer}}\ and\ \bibinfo {author} {\bibfnamefont {R.~L.}\ \bibnamefont
  {Graham}},\ }\href {\doibase http://dx.doi.org/10.1016/j.jpcs.2006.05.015}
  {\bibfield  {journal} {\bibinfo  {journal} {J. Phys. Chem. Solids}\ }\textbf
  {\bibinfo {volume} {67}},\ \bibinfo {pages} {2136 } (\bibinfo {year}
  {2006})}\BibitemShut {NoStop}%
\bibitem [{\citenamefont {Militzer}(2006)}]{PhysRevLett.97.175501}%
  \BibitemOpen
  \bibfield  {author} {\bibinfo {author} {\bibfnamefont {B.}~\bibnamefont
  {Militzer}},\ }\href {\doibase 10.1103/PhysRevLett.97.175501} {\bibfield
  {journal} {\bibinfo  {journal} {Phys. Rev. Lett.}\ }\textbf {\bibinfo
  {volume} {97}},\ \bibinfo {pages} {175501} (\bibinfo {year}
  {2006})}\BibitemShut {NoStop}%
\bibitem [{\citenamefont {Militzer}(2005)}]{Militzer2005}%
  \BibitemOpen
  \bibfield  {author} {\bibinfo {author} {\bibfnamefont {B.}~\bibnamefont
  {Militzer}},\ }\href {\doibase 10.1007/s10909-005-5485-8} {\bibfield
  {journal} {\bibinfo  {journal} {J. Low Temp. Phys.}\ }\textbf {\bibinfo
  {volume} {139}},\ \bibinfo {pages} {739} (\bibinfo {year}
  {2005})}\BibitemShut {NoStop}%
\bibitem [{\citenamefont {Militzer}\ and\ \citenamefont
  {Pollock}(2005)}]{PhysRevB.71.134303}%
  \BibitemOpen
  \bibfield  {author} {\bibinfo {author} {\bibfnamefont {B.}~\bibnamefont
  {Militzer}}\ and\ \bibinfo {author} {\bibfnamefont {E.~L.}\ \bibnamefont
  {Pollock}},\ }\href {\doibase 10.1103/PhysRevB.71.134303} {\bibfield
  {journal} {\bibinfo  {journal} {Phys. Rev. B}\ }\textbf {\bibinfo {volume}
  {71}},\ \bibinfo {pages} {134303} (\bibinfo {year} {2005})}\BibitemShut
  {NoStop}%
\bibitem [{\citenamefont {Pollock}\ and\ \citenamefont
  {Militzer}(2004)}]{PhysRevLett.92.021101}%
  \BibitemOpen
  \bibfield  {author} {\bibinfo {author} {\bibfnamefont {E.~L.}\ \bibnamefont
  {Pollock}}\ and\ \bibinfo {author} {\bibfnamefont {B.}~\bibnamefont
  {Militzer}},\ }\href {\doibase 10.1103/PhysRevLett.92.021101} {\bibfield
  {journal} {\bibinfo  {journal} {Phys. Rev. Lett.}\ }\textbf {\bibinfo
  {volume} {92}},\ \bibinfo {pages} {021101} (\bibinfo {year}
  {2004})}\BibitemShut {NoStop}%
\bibitem [{\citenamefont {Ceperley}(1991)}]{Ceperley1991}%
  \BibitemOpen
  \bibfield  {author} {\bibinfo {author} {\bibfnamefont {D.~M.}\ \bibnamefont
  {Ceperley}},\ }\href@noop {} {\bibfield  {journal} {\bibinfo  {journal} {J.
  Stat. Phys.}\ }\textbf {\bibinfo {volume} {63}},\ \bibinfo {pages} {1237}
  (\bibinfo {year} {1991})}\BibitemShut {NoStop}%
\bibitem [{Kha()}]{Khairallah2011}%
  \BibitemOpen
  \href@noop {} {}\bibinfo {note} {S. A. Khairallah, J. Shumway, and E. W.
  Draeger, arXiv:1108.1711}\BibitemShut {NoStop}%
\bibitem [{\citenamefont {Militzer}(2000)}]{militzerphd}%
  \BibitemOpen
  \bibfield  {author} {\bibinfo {author} {\bibfnamefont {B.}~\bibnamefont
  {Militzer}},\ }\href@noop {} {\bibinfo {type} {{Ph.D. Thesis}}},\ \bibinfo
  {school} {University of Illinois at Urbana-Champaign} (\bibinfo {year}
  {2000})\BibitemShut {NoStop}%
\bibitem [{\citenamefont {Militzer}(2016)}]{pdm}%
  \BibitemOpen
  \bibfield  {author} {\bibinfo {author} {\bibfnamefont {B.}~\bibnamefont
  {Militzer}},\ }\href@noop {} {\bibfield  {journal} {\bibinfo  {journal}
  {Comput. Phys. Commun.}\ }\textbf {\bibinfo {volume} {204}},\ \bibinfo
  {pages} {88} (\bibinfo {year} {2016})}\BibitemShut {NoStop}%
\bibitem [{\citenamefont {Kresse}\ and\ \citenamefont
  {Furthm{\"u}ller}(1996)}]{kresse96b}%
  \BibitemOpen
  \bibfield  {author} {\bibinfo {author} {\bibfnamefont {G.}~\bibnamefont
  {Kresse}}\ and\ \bibinfo {author} {\bibfnamefont {J.}~\bibnamefont
  {Furthm{\"u}ller}},\ }\href@noop {} {\bibfield  {journal} {\bibinfo
  {journal} {Phys. Rev. B}\ }\textbf {\bibinfo {volume} {54}},\ \bibinfo
  {pages} {11169} (\bibinfo {year} {1996})}\BibitemShut {NoStop}%
\bibitem [{\citenamefont {Perdew}\ and\ \citenamefont
  {Zunger}(1981)}]{Perdew81}%
  \BibitemOpen
  \bibfield  {author} {\bibinfo {author} {\bibfnamefont {J.~P.}\ \bibnamefont
  {Perdew}}\ and\ \bibinfo {author} {\bibfnamefont {A.}~\bibnamefont
  {Zunger}},\ }\href@noop {} {\bibfield  {journal} {\bibinfo  {journal} {Phys.
  Rev. B}\ }\textbf {\bibinfo {volume} {23}},\ \bibinfo {pages} {5048}
  (\bibinfo {year} {1981})}\BibitemShut {NoStop}%
\bibitem [{\citenamefont {Ceperley}\ and\ \citenamefont
  {Alder}(1980)}]{Ceperley1980}%
  \BibitemOpen
  \bibfield  {author} {\bibinfo {author} {\bibfnamefont {D.~M.}\ \bibnamefont
  {Ceperley}}\ and\ \bibinfo {author} {\bibfnamefont {B.~J.}\ \bibnamefont
  {Alder}},\ }\href@noop {} {\bibfield  {journal} {\bibinfo  {journal} {Phys.
  Rev. Lett.}\ }\textbf {\bibinfo {volume} {45}},\ \bibinfo {pages} {566}
  (\bibinfo {year} {1980})}\BibitemShut {NoStop}%
\bibitem [{\citenamefont {Bl\"ochl}, \citenamefont {Jepsen},\ and\
  \citenamefont {Andersen}(1994)}]{Blochl1994}%
  \BibitemOpen
  \bibfield  {author} {\bibinfo {author} {\bibfnamefont {P.~E.}\ \bibnamefont
  {Bl\"ochl}}, \bibinfo {author} {\bibfnamefont {O.}~\bibnamefont {Jepsen}}, \
  and\ \bibinfo {author} {\bibfnamefont {O.~K.}\ \bibnamefont {Andersen}},\
  }\href@noop {} {\bibfield  {journal} {\bibinfo  {journal} {Phys. Rev. B}\
  }\textbf {\bibinfo {volume} {49}},\ \bibinfo {pages} {16223} (\bibinfo {year}
  {1994})}\BibitemShut {NoStop}%
\bibitem [{com()}]{comment1}%
  \BibitemOpen
  \href@noop {} {}\bibinfo {note} {{We checked the inter-atomic
  distance and found that the 1.45-Bohr core radius and $\sim$1.2-Bohr 
  projector cutoff radius defined in the pseudopotential are small enough to avoid core
  overlap at all densities up to 9.67 g/cm$^3$ (10 $\rho_\text{ambient}$). At
  11.6 g/cm$^3$ (12 $\rho_\text{ambient}$), the cores of the nearest neighbors
  overlap by less than 0.2 Bohr, while the projectors do not overlap and the
  EOS data from DFT-MD still agree very well with PIMC values at 1 and 2
  million K (within 1.6 Ha/atom in energy and 2\% in pressure). Moreover,
  removing data at this density from the EOS table does not affect the Hugoniot
  curves nor does it matter for our discussion of the shock Hugoniot
  curves.}}\BibitemShut {Stop}%
\bibitem [{opi()}]{opium}%
  \BibitemOpen
  \href@noop {} {}\bibinfo {note} {{http://opium.sourceforge.net}}\BibitemShut
  {NoStop}%
\bibitem [{\citenamefont {Bahcall}\ and\ \citenamefont
  {Pinsonneault}(2004)}]{PhysRevLett.92.121301}%
  \BibitemOpen
  \bibfield  {author} {\bibinfo {author} {\bibfnamefont {J.~N.}\ \bibnamefont
  {Bahcall}}\ and\ \bibinfo {author} {\bibfnamefont {M.~H.}\ \bibnamefont
  {Pinsonneault}},\ }\href {\doibase 10.1103/PhysRevLett.92.121301} {\bibfield
  {journal} {\bibinfo  {journal} {Phys. Rev. Lett.}\ }\textbf {\bibinfo
  {volume} {92}},\ \bibinfo {pages} {121301} (\bibinfo {year}
  {2004})}\BibitemShut {NoStop}%
\bibitem [{CGp()}]{CGpt}%
  \BibitemOpen
  \href@noop {} {}\bibinfo {note} {Private communication with Cyril Georgy,
  University of Keele (United Kingdom).}\BibitemShut {Stop}%
\bibitem [{\citenamefont {Perdew}, \citenamefont {Burke},\ and\ \citenamefont
  {Ernzerhof}(1996)}]{Perdew96}%
  \BibitemOpen
  \bibfield  {author} {\bibinfo {author} {\bibfnamefont {J.~P.}\ \bibnamefont
  {Perdew}}, \bibinfo {author} {\bibfnamefont {K.}~\bibnamefont {Burke}}, \
  and\ \bibinfo {author} {\bibfnamefont {M.}~\bibnamefont {Ernzerhof}},\
  }\href@noop {} {\bibfield  {journal} {\bibinfo  {journal} {Phys. Rev. Lett.}\
  }\textbf {\bibinfo {volume} {77}},\ \bibinfo {pages} {3865} (\bibinfo {year}
  {1996})}\BibitemShut {NoStop}%
\bibitem [{gam()}]{gamess}%
  \BibitemOpen
  \href@noop {} {}\bibinfo {note}
  {{http://www.msg.ameslab.gov/gamess/}}\BibitemShut {NoStop}%
\bibitem [{Note1()}]{Note1}%
  \BibitemOpen
  \bibinfo {note} {Here, we include the kinetic energy of the nuclei when
  calculating the total internal energy of the ideal Fermi electron gas model,
  but have not considered the charge and interaction of the
  nuclei.}\BibitemShut {Stop}%
\bibitem [{\citenamefont {Ahrens}(2003)}]{Ahrens2003}%
  \BibitemOpen
  \bibfield  {author} {\bibinfo {author} {\bibfnamefont {T.~J.}\ \bibnamefont
  {Ahrens}},\ }in\ \href {\doibase 10.1007/978-1-4020-4423-6_293} {\emph
  {\bibinfo {booktitle} {Encyclopedia of Geomagnetism and Paleomagnetism}}},\
  \bibinfo {editor} {edited by\ \bibinfo {editor} {\bibfnamefont
  {D.}~\bibnamefont {Gubbins}}\ and\ \bibinfo {editor} {\bibfnamefont
  {E.}~\bibnamefont {Herrero-Bervera}}}\ (\bibinfo  {publisher} {Springer
  Netherlands},\ \bibinfo {address} {Dordrecht},\ \bibinfo {year} {2003})\ pp.\
  \bibinfo {pages} {912--920}\BibitemShut {NoStop}%
\bibitem [{\citenamefont {Lyon}\ and\ \citenamefont {Johnson}(1992)}]{sesame}%
  \BibitemOpen
  \bibinfo {editor} {\bibfnamefont {S.~P.}\ \bibnamefont {Lyon}}\ and\ \bibinfo
  {editor} {\bibfnamefont {J.~D.}\ \bibnamefont {Johnson}},\ eds.,\ \href@noop
  {} {\emph {\bibinfo {title} {SESAME: The Los Alamos National Laboratory
  Equation of State Database}}}\ (\bibinfo  {publisher} {Group T-1, Report No.
  LA-UR-92-3407},\ \bibinfo {year} {1992})\BibitemShut {NoStop}%
\bibitem [{\citenamefont {More}\ \emph {et~al.}(1988)\citenamefont {More},
  \citenamefont {Warren}, \citenamefont {Young},\ and\ \citenamefont
  {Zimmerman}}]{leos1}%
  \BibitemOpen
  \bibfield  {author} {\bibinfo {author} {\bibfnamefont {R.~M.}\ \bibnamefont
  {More}}, \bibinfo {author} {\bibfnamefont {K.~H.}\ \bibnamefont {Warren}},
  \bibinfo {author} {\bibfnamefont {D.~A.}\ \bibnamefont {Young}}, \ and\
  \bibinfo {author} {\bibfnamefont {G.~B.}\ \bibnamefont {Zimmerman}},\ }\href
  {\doibase http://dx.doi.org/10.1063/1.866963} {\bibfield  {journal} {\bibinfo
   {journal} {Phys. Fluids}\ }\textbf {\bibinfo {volume} {31}},\ \bibinfo
  {pages} {3059} (\bibinfo {year} {1988})}\BibitemShut {NoStop}%
\bibitem [{\citenamefont {Young}\ and\ \citenamefont {Corey}(1995)}]{leos2}%
  \BibitemOpen
  \bibfield  {author} {\bibinfo {author} {\bibfnamefont {D.~A.}\ \bibnamefont
  {Young}}\ and\ \bibinfo {author} {\bibfnamefont {E.~M.}\ \bibnamefont
  {Corey}},\ }\href {\doibase http://dx.doi.org/10.1063/1.359955} {\bibfield
  {journal} {\bibinfo  {journal} {J. Appl. Phys.}\ }\textbf {\bibinfo {volume}
  {78}},\ \bibinfo {pages} {3748} (\bibinfo {year} {1995})}\BibitemShut
  {NoStop}%
\bibitem [{\citenamefont {Wilson}\ \emph {et~al.}(2006)\citenamefont {Wilson},
  \citenamefont {Sonnad}, \citenamefont {Sterne},\ and\ \citenamefont
  {Isaacs}}]{Purgatorio2006}%
  \BibitemOpen
  \bibfield  {author} {\bibinfo {author} {\bibfnamefont {B.}~\bibnamefont
  {Wilson}}, \bibinfo {author} {\bibfnamefont {V.}~\bibnamefont {Sonnad}},
  \bibinfo {author} {\bibfnamefont {P.}~\bibnamefont {Sterne}}, \ and\ \bibinfo
  {author} {\bibfnamefont {W.}~\bibnamefont {Isaacs}},\ }\href@noop {}
  {\bibfield  {journal} {\bibinfo  {journal} {J. Quant. Spectrosc. Radiat.
  Transfer}\ }\textbf {\bibinfo {volume} {99}},\ \bibinfo {pages} {658}
  (\bibinfo {year} {2006})}\BibitemShut {NoStop}%
\bibitem [{Whi()}]{WhitleyLLNL}%
  \BibitemOpen
  \href@noop {} {}\bibinfo {note} {H. D. Whitley and C. J. Wu, Lawrence
  Livermore National Laboratory Report No. LLNL-TR-705163 (2016)}\BibitemShut
  {NoStop}%
\bibitem [{\citenamefont {Ozaki}\ \emph {et~al.}(2016)\citenamefont {Ozaki},
  \citenamefont {Nellis}, \citenamefont {Mashimo}, \citenamefont {Ramzan},
  \citenamefont {Ahuja}, \citenamefont {Kaewmaraya}, \citenamefont {Kimura},
  \citenamefont {Knudson}, \citenamefont {Miyanishi}, \citenamefont {Sakawa},
  \citenamefont {Sano},\ and\ \citenamefont {Kodama}}]{Ozaki2016}%
  \BibitemOpen
  \bibfield  {author} {\bibinfo {author} {\bibfnamefont {N.}~\bibnamefont
  {Ozaki}}, \bibinfo {author} {\bibfnamefont {W.~J.}\ \bibnamefont {Nellis}},
  \bibinfo {author} {\bibfnamefont {T.}~\bibnamefont {Mashimo}}, \bibinfo
  {author} {\bibfnamefont {M.}~\bibnamefont {Ramzan}}, \bibinfo {author}
  {\bibfnamefont {R.}~\bibnamefont {Ahuja}}, \bibinfo {author} {\bibfnamefont
  {T.}~\bibnamefont {Kaewmaraya}}, \bibinfo {author} {\bibfnamefont
  {T.}~\bibnamefont {Kimura}}, \bibinfo {author} {\bibfnamefont
  {M.}~\bibnamefont {Knudson}}, \bibinfo {author} {\bibfnamefont
  {K.}~\bibnamefont {Miyanishi}}, \bibinfo {author} {\bibfnamefont
  {Y.}~\bibnamefont {Sakawa}}, \bibinfo {author} {\bibfnamefont
  {T.}~\bibnamefont {Sano}}, \ and\ \bibinfo {author} {\bibfnamefont
  {R.}~\bibnamefont {Kodama}},\ }\href {http://dx.doi.org/10.1038/srep26000
  http://10.1038/srep26000} {\bibfield  {journal} {\bibinfo  {journal} {Sci.
  Rep.}\ }\textbf {\bibinfo {volume} {6}},\ \bibinfo {pages} {26000} (\bibinfo
  {year} {2016})}\BibitemShut {NoStop}%
\end{thebibliography}

%

\end{document}